\documentclass[a4paper,11pt]{article}

\usepackage{jheppub} 

\usepackage{graphicx}

\usepackage{amsmath}
\usepackage[utf8]{inputenc}

\usepackage{amssymb}
\usepackage{amsfonts}
\usepackage{amsbsy}
\usepackage{url}
\usepackage{enumerate}
\usepackage{caption}
\usepackage{subcaption}
\usepackage{color}
\usepackage{ulem}
\usepackage{bm}
\usepackage{comment}
\usepackage{braket}
\usepackage{tikz}
\usepackage{wrapfig}
\usepackage{physics}

\newcommand{\be}{\begin{eqnarray}}
\newcommand{\ee}{\end{eqnarray}}
\newcommand{\p}{\partial}

\makeatletter
\@addtoreset{equation}{section}
\makeatother

\newcommand{\refer}[1]{(\ref{#1})}
\newcommand{\bee}{\begin{equation}}
\newcommand{\eee}{(\end{equation})}

\newcommand{\expon}{\mathrm{e}}

\newcommand\rsout{\bgroup\markoverwith{\textcolor{red}{\rule[0.5ex]{2pt}{0.4pt}}}\ULon}

\title{
Massless monopole-string-domain wall fermions and 
polyhedral vacuum fermions
}

\author[a,b,c]{Minoru Eto}
\emailAdd{meto[at]sci.kj.yamagata-u.ac.jp}
\affiliation[a]{Department of Physics, Yamagata University, 
Kojirakawa-machi 1-4-12, Yamagata, Yamagata 990-8560, Japan}

\affiliation[b]{Research and Education Center for Natural Sciences, Keio University, 4-1-1 Hiyoshi, Yokohama, Kanagawa 223-8521, Japan}

\affiliation[c]{International Institute for Sustainability with Knotted Chiral Meta Matter(WPI-SKCM$^2$), Hiroshima University, 1-3-2 Kagamiyama, Higashi-Hiroshima, Hiroshima 739-8511, Japan}

\author[a]{and Yuito Suzuki}
\emailAdd{s231862d[at]st.yamagata-u.ac.jp}

\abstract{
Fermion zero modes of  Bogomol'nyi-Prasad-Sommerfield monopole-string-domain wall composites in three spatial dimensions are studied. We analytically solve the Dirac equation and prove the existence of one fermion zero mode. 
Depending on mass parameters of bosons/fermions in the model, the zero modes are localized either on the monopoles, strings or domain walls, which we call monopole-string-domain wall fermions. We also show that in special cases, the zero modes can be confined within a finite vacuum region in the shape of an arbitrary convex polyhedron, which we call the polyhedral vacuum fermions. Furthermore, we show that fermionic superconducting currents do not generally flow on the host solitons except for the cases that the soliton network consists only of strings and domain walls and has translational symmetry about a spatial axis.
}
\preprint{YGHP-25-02}

\begin{document}
\maketitle


\section{Introduction}

Topological solitons play a crucial role in various areas of physics, and therefore they have been intensively studied in  particle physics, nuclear physics, cosmology, and condensed matter physics.
The universality of topological solitons stems from a phenomenon called spontaneous symmetry breaking (SSB), a simple yet powerful idea that does not depend on the detailed structure of 
systems.
When a topologically nontrivial order parameter space appears through SSB in a model, different types of topological solitons, depending of the topological characteristics of the SSB, such as domain walls, strings\cite{Abrikosov:1956sx,Nielsen:1973cs}, monopoles \cite{tHooft:1974kcl,Polyakov:1974ek}, and instantons \cite{Belavin:1975fg}, may emerge as topologically nontrivial excitations.

There are two interesting generalization on the topological solitons. 
The one is that several different species of topological solitons would coexist and form composite states in a certain situation.
They can be generated when multiple SSBs occur in several stages or simultaneously \cite{Preskill:1992ck,Eto:2023gfn}.
Although there are lots of researches on composite solitons recently, they are still relatively underexplored compared to the long 
history of research on single type of solitons. 
The other is  localization of massless fermions on the topological solitons of a single species. 
The massless fermions exhibit the curious property that the host soliton carries a fractional fermion number \cite{Jackiw:1975fn, Jackiw:1981ee}.
Furthermore, they make host solitons superconducting \cite{WITTEN1985557}.

One of the fields in which both the composite solitons and massless fermions are relevant is the axion cosmology  \cite{Peccei:1977hh,Weinberg:1977ma,Wilczek:1977pj}.
Axion is a hypothetical Nambu-Goldstone particle associated with SSB of a global \(U(1)_\mathrm{PQ}\) symmetry which is introduced to solve the strong CP problem by Peccei-Quinn (PQ) mechanism.
It is also important as a candidate of a viable cold dark matter \cite{Preskill:1982cy,Abbott:1982af,Dine:1982ah}.
When the \(U(1)_\mathrm{PQ}\) symmetry is broken in the early Universe, a large number of cosmic strings are formed by the Kibble-Zurek mechanism \cite{Kibble:1976sj,Zurek:1985qw}.
These strings collide and form loops, which subsequently shrink and decay, resulting in reducing the number of strings.
Lifetime of the cosmic strings is important because it overcloses the universe if it has a lifetime of cosmological timescale, 
and also it affects axion relic abundance through evolution of the universe \cite{Davis:1986xc}. 
The superconductivity is one of the possible effect which can significantly affect the dynamics cosmic strings \cite{Lazarides:1984zq,Chudnovsky:1986hc,Ostriker:1986xc,Copeland:1987th,Nielsen:1987fy,Davis:1988jp,CALLAN1985427}.
If the fermions coupled to the string carry a non-zero electric charge, they can travel along the string without resistance, forming a superconducting current. The current can prevent the loops from shrinking, making the superconducting loops, called vortons \cite{Carter:1993wu,Brandenberger:1996zp,Martins:1998gb,Martins:1998th,Carter:1999an}, potentially long-lived.
If the vortons have a sufficiently long lifetime, they dominate the energy density of the universe. Recently, the superconducting axion cosmic string  has attracted renewed interest \cite{Fukuda:2020kym,Abe:2020ure,Agrawal:2020euj,Ibe:2021ctf}.
The axion string in the KSVZ axion model \cite{Kim:1979if,Shifman:1979if} has been shown to be a chiral superconductor in Ref.~\cite{Fukuda:2020kym}, and several interesting phenomena related to chiral superconductivity have been found:
The decay of superconducting current due to a current leakage processes,
increase in the number of strings per Hubble volume due to interaction between strings carrying enhanced currents in the primordial magnetic field and the surrounding plasma, and shift of the QCD axion relic abundance window in some parts of the parameter space.

Most of these studies have considered the fermionic superconductors of the cosmic strings. 
However, the \(U(1)_\mathrm{PQ}\) symmetry is broken down to \(Z_N\) symmetry at the QCD phase transition, so that the 
axion cosmic strings are inevitably attached by \(N\) domain walls.
This is cosmologically important because the axion dark matter density is produced by
the decay of the string-domain wall network at the QCD phase transition \cite{Hiramatsu:2010yn,Hiramatsu:2012gg,Gorghetto:2018myk,Gorghetto:2020qws}.
A natural question then arises: whether fermionic superconductivity remains at the strings, or whether it appears only at the domain walls, or even throughout the entire string-domain wall composite.
If massless fermions are localized on domain walls or string-domain wall composites rather than on strings, this would significantly affect on the axion relic abundance.
With this question in mind, in our previous work \cite{Eto:2023orr}, we investigated fermion zero modes on a generic string-domain wall composite solitons in an axion(-like) model and an Abelian Higgs model inspired by \(\mathcal{N}=2\) supersymmetry.
We showed the existence of fermion zero modes, and we found that they can be localized on either a string, a domain wall, or in one of the vacua.\footnote{Note that this is quite different from the fermion zero modes localized across the domain wall junction (the string and domain wall) in the genuine ${\cal N}=1$ supersymmetric Abelian-Higgs model \cite{Ito:2000zf}, due to the special property that the fermion zero mode is a superpartner of the translational zero mode.}
Especially, we obtained analytic solutions of the fermion zero mode functions in the \(\mathcal{N}=2\) SUSY inspired Abelian-Higgs model.
We also showed that the superconducting currents flowing along the strings, domain walls, or the vacuum are chiral currents~\cite{Eto:2023orr}.

In this paper, we continue the previous studies about massless fermions on the soliton composites.
The composite solitons considered in \cite{Eto:2023orr} are not the most general ones, since we have only considered networks of strings and domain walls that are translationally symmetric along a spatial axis (e.g., the \(z\) axis).
In contrast, in this paper we consider 
more general configurations of composite solitons expanding in three dimensions. The three-dimensional network of solitons consists not only of strings and domain walls, but also of point-like solitons that are global monopoles. Therefore, the aim of this paper is to examine the existence of fermion zero modes under a three-dimensional network of solitons consisting of monopoles, strings, and domain walls, and to clarify in which part of the network the zero modes are localized, if they exist.
To this end, we will work in a particular model that has an advantage in its ease of handling three-dimensional soliton networks.
The bosonic part of the Lagrangian is the \(\mathcal{N}=2\) supersymmetry inspired Abelian-Higgs model investigated in Refs.~\cite{Eto:2020vjm,Eto:2020cys}. In this model, a network of composite solitons appears as a Bogomol'nyi-Prasad-Sommerfield (BPS) state satisfying a first-order differential equation and is thereby well controlled by the established moduli matrix formalism \cite{Eto:2005cp,Eto:2005fm}.
For fermionic part of the Lagrangian, we will consider a simple Yukawa coupling between the order parameter scalar fields $\varphi_m$ $(m=1,2,3)$
and four component Dirac fermion $\Psi_a$ ($a=1,2$).
We will analytically solve the Dirac equation and obtain a fermionic zero mode under the background of the BPS monopole-string-domain wall composites. Using the analytical solutions, we show that the zero mode can be localized either
on the monopole, the string or the domain wall, and explain that where it actually appears depends on the mass parameters of the model.
We also show that massless fermions can be confined within a finite vacuum region in the form of an arbitrary convex  polyhedron, such as the Platonic and Archimedean polyhedron. Furthermore, we show that vector supercurrents, but not chiral supercurrents, flow only in networks with translational symmetry along the spatial axis.

The organization of this paper is as follows.
In Section 2, we review the three dimensional BPS composite solitons in the SUSY-inspired model.
In Section 3, we analytically solve the Dirac equation and show zero-mode solutions.
We then investigate their localization properties and supercurrents.
Section 4 is for conclusions and discussions of future directions.

\section{Brief review on global monopole-string-domain wall composites}
\label{sec:GMSW}

\subsection{The SUSY inspired model}
\label{sec:sec:GMSW_1}

In this section we will review solitons that are composites of three different types: global monopoles, strings and domain walls.
We will study a non-supersymmetric model which admit these composite solitons 
as Bogomol'nyi-Prasad-Sommerfield (BPS) states \cite{Eto:2020vjm,Eto:2020cys}.

Let us begin with the bosonic Lagrangian inspired by the \(3+1\) dimensional \(\mathcal{N}=2\) supersymmetric 
Abelian-Higgs model with $N_{\rm F}$ Higgs fields:
\begin{align}
\mathcal{L}&=-\frac{1}{4e^2}F_{\mu\nu}F^{\mu\nu}+\frac{1}{2e^2}\sum_{m=1}^3\partial_\mu\varphi_m\partial^\mu\varphi_m+\abs{D_\mu H}^2-V \\
V&=\frac{e^2}{2}\qty(v^2-\abs{H}^2)^2+\sum_{m=1}^3\abs{\varphi_m H-HM_m}^2.
\label{eq:Lagrangian_boson}
\end{align}
Here, \(A_\mu\) is the \(U(1)\) gauge field, \(\varphi_m~(m=1,2,3)\) are the real scalar fields, and \(H\) is the \(N_\mathrm{F}\) components row vector.
We denote its components as \(H^A (A=1,2,\cdots,N_\mathrm{F})\). The covariant derivative is given by $D_\mu H = \p_\mu H + i A_\mu H$.
The Lagrangian does not respect to SUSY:
In the original ${\cal N} = 2$ SUSY Abelian-Higgs model \cite{Isozumi:2003rp,Isozumi:2003uh,Isozumi:2004jc,Isozumi:2004va,Isozumi:2004vg,Eto:2004rz,Eto:2005cp,Eto:2005sw,Sakai:2006prl,Eto:2005fm,Eto:2006pg}, $A_\mu$ and $\varphi_{1,2}$ appear as the bosonic component of the vector multiplet, and $H$ forming $SU(2)_R$ doublet together with another complex scalar field $\tilde H$ is a member of the hypermultiplet. In our model we add the third real scalar $\varphi_3$ and ignore $\tilde H$,
so that the Lagrangian does not respect SUSY.
Furthermore, all the fermions are discarded (we include fermions in Sec.~\ref{sec:MSDWF}).
Here \(e\) is the \(U(1)\) gauge coupling and \(M_m~(m=1,2,3)\) is the diagonal mass matrices of the size $N_{\rm F} \times N_{\rm F}$ for \(H\) as
\begin{gather}
M_m=\mathrm{diag}(m_{m,1},m_{m,2},\cdots,m_{m,N_\mathrm{F}}),\quad(m=1,2,3).
\end{gather}
For later convenience, we introduce the \(N_\mathrm{F}\) mass vectors defined by
\begin{gather}
\vec{m}_{A}=
\begin{pmatrix}
m_{1,A} \\
m_{2,A} \\
m_{3,A}
\end{pmatrix}
,\quad (A=1,2,\cdots,N_\mathrm{F}).
\end{gather}

We assume thorough out this paper that the mass vectors are not fully degenerate, namely \(\vec m_A \neq \vec m_B\) if \(A\neq B\).
Then there are \(N_\mathrm{F}\) discrete vacua satisfying \(V=0\).
The \(A\)-th vacuum denoted by $\left<\cdots\right>_A$ is given by
\begin{gather}
\ev{H^B}_A=v\delta^B_A,\quad\ev{\varphi_m}_A=m_{m,A}.
\label{eq:vacuum}
\end{gather}
Hence, the $A$-th vacuum corresponds to the point \(\vec \varphi= \vec m_A\) in the three dimensional \(\varphi_m\) space.

Note that if the mass matrix is proportional to the identity matrix as $M_m = m_m {\bf 1}_{N_{\rm F}}$, the model admits the $SU(N_{\rm F})$ flavor symmetry $H \to H U$ ($U \in SU(N_{\rm F})$). But it is reduces to the Cartan subalgebra $U(1)^{N_{\rm F}-1}$ for the generic mass matrix.
Note also that if the mass matrix is zero $M_m = 0$, then the model has the additional global symmetry $SU(2)$: $\vec \varphi \to G \vec \varphi$ ($G \in SU(2)$).

\subsection{The BPS equations and the moduli matrix}

In the following, we consider static configurations. Then, we set \(\partial_0=0\), and we work in the temporal gauge \(A_0=0\).
It is possible to perform the Bogomol'nyi completion for the energy density, which is the reason why we study this non-SUSY model, as \cite{Eto:2020vjm,Eto:2020cys}
\begin{align}
\mathcal{E}=&\frac{1}{2e^2}\sum_{m>n}\qty{F_{mn}^2+(\partial_m\varphi_n-s_ms_n\partial_n\varphi_m)^2}+\frac{1}{2e^2}\qty{\sum_{m}s_m\partial_m\varphi_m-e^2\qty(v^2-\abs{H}^2)}^2 \nonumber \\
&+\sum_m\abs{D_mH+s_m\qty(\varphi_mH-HM_m)}^2+v^2\sum_ms_m\mathcal{W}_m
- \frac{1}{e^2}\sum_{m>n}s_ms_n\mathcal{S}_{mn}+\sum_ms_m\partial_m\mathcal{J}_m \nonumber \\
\geq&~v^2\sum_ms_m\mathcal{W}_m - \frac{1}{e^2}\sum_{m>n}s_ms_n\mathcal{S}_{mn}+\sum_ms_m\partial_m\mathcal{J}_m,
\label{eq:energydensity}
\end{align}
where are \(\mathcal{W}_m$ and $\mathcal{S}_{mn}\) are the topological charge densities for the domain walls and strings, respectively, defined by
\begin{align}
\mathcal{W}_m&=\partial_m\varphi_m, \\
\mathcal{S}_{mn}&=\partial_m\varphi_m\partial_n\varphi_n-\partial_m\varphi_n\partial_n\varphi_m.
\end{align}
The perfect squaring always allows us to choose signs as $a^2 + b^2 = (a \pm b)^2 \mp 2 ab$. 
In the above squaring procedure above, this freedom is represented by the symbol $s_m = \pm 1$. 
Hence, there are 8 possible combinations of choosing the sign of $(s_1,s_2,s_3)$.
We also define ${\cal J}_m$ by
\begin{align}
\mathcal{J}_m&=-(\varphi_mH-HM_m)H^\dag,
\end{align}
which is not important since it vanishes at any vacuum.

The energy  density is bounded from below, and it is saturated when the following BPS equations are satisfied
\begin{gather}
F_{mn}=0,\quad s_n\partial_m\varphi_n-s_m\partial_n\varphi_m=0,\quad\qty(s_mD_m+\varphi_m-m_{m,A})H^A=0, \label{eq:BPS1} \\
\sum_m s_m\partial_m\varphi_m=e^2\left(v^2-|H|^2\right), \label{eq:BPS2}
\end{gather}
where the sums over $m$, $n$, and $A$ are not taken unless explicitly stated.

The topological charge of the domain wall is given by integrating ${\cal W}_m$ as
\be
W_m = \int^\infty_{-\infty} dx^m~\p_m \varphi_m
= \varphi_m\big|_{x^m \to \infty} - \varphi_m\big|_{x^m \to -\infty}\,,
\label{eq:domain_wall}
\ee
where no sum is taken over $m$. Note that $s_m W_m$ is always positive regardless of the choice of $s_m$, and the tension (energy per unit area) of the domain wall measured along the normal direction to the domain wall interpolating the $A$-th and the $B$-th vacua is given by
\be
\left| \vec W \right| = v^2 \left| \vec m_A - \vec m_B\right|\,.
\ee
Similarly, the topological charge for the strings is given by
\be
S_{mn} = \int dx^mdx^n~ {\cal S}_{mn}.
\label{eq:string}
\ee
Since ${\cal S}_{mn}$ is a Jacobian of the map from $x^m$-$x^n$ plane to $\varphi_m$-$\varphi_n$ plane, $S_{mn}$ measures area of image of the map. It can be either positive or negative but the combination $s_m s_n S_{mn}$ is always positive. Thus, the string tension always negatively contributes to total energy, and therefore it can be understood as a binding energy of three domain walls which meet at a string \cite{Oda:1999az}.

We now realize that the topological objects in our model have certain geometrical meaning in the $\varphi_m$ space.
The vacua correspond to the vertices $\vec\varphi = \vec m_A$ as Eq.~(\ref{eq:vacuum}), the domain walls are related to the edges between two vertices by $W_m$ in Eq.~(\ref{eq:domain_wall}),
and the strings correspond to the faces with three vertices by $S_{mn}$ in Eq.~(\ref{eq:string}).
These correspondence is naturally extended to one dimension higher object (body of a polyhedron) with four vertices, and a natural extension
of lower dimensional topological charge densities is given by
\begin{gather}
\mathcal{M}= \sum_{l,m,n} \epsilon^{lmn}\partial_l\varphi_1\partial_m\varphi_2\partial_n\varphi_3.
\label{eq:M_density}
\end{gather}
This is a Jacobian of the map from $(x^1,x^2,x^3)$ to $(\varphi_1,\varphi_2,\varphi_3)$, and its integral
\be
M = \int d^3x~ {\cal M}\,,
\label{eq:M}
\ee
measures the volume of the polyhedron. Physically, it is a point-like object, and we call it a monopole, and indeed it appears
at a junction of four strings meet. These will be explained better below through concrete examples. See also the appendix \ref{sec:appA} for
a familiar spherically symmetric global monopole.

Note that the similar Bogomol'nyi completion in the original ${\cal N}=2$ SUSY Abelian-Higgs model 
without the third field $\varphi^3$ have been investigated long time ago \cite{Eto:2005cp,Eto:2005fm}.
The corresponding solitons are composites of the strings and domain walls only, and they are essentially two dimensional solitons
because they trivially extend along one spatial direction, say $x^3$.

Now, we understand the BPS equations (\ref{eq:BPS1}) and (\ref{eq:BPS2}) admits composite of the monopoles, strings, and domain walls. 
Explicit solutions of the BPS equations are constructed by the moduli matrix formalism method\cite{Isozumi:2004jc,Isozumi:2004va,Isozumi:2004vg,Eto:2005cp,Eto:2005fm,Eto:2006pg,Eto:2020vjm,Eto:2020cys}. 
Let us first rewrite $H$, $\varphi_m$, and $A_m$ as
\begin{gather}
H^A=vS^{-1}H_0^A \exp\left(\sum_m s_mm_{m,A}x^m\right)
,\quad A_m-is_m\varphi_m=-iS^{-1}\partial_mS,
\label{eq:solofmodu}
\end{gather}
where \(S\) is a non-zero complex scalar function and \(H_0\) is an \(1\times N_\mathrm{F}\) arbitrary constant complex matrix called the moduli matrix.
Then the BPS equations in Eq.~(\ref{eq:BPS1}) are automatically satisfied.
We are left with Eq. (\ref{eq:BPS2}) which is rewritten as 
\begin{gather}
\frac{1}{2e^2v^2}\partial^2\log\Omega=1-\Omega_0\Omega^{-1},\quad\Omega=|S|^2,
\label{eq:master}
\end{gather}
with
\begin{gather}
\Omega_0=\sum_{A}|H^A_0|^2
\exp\left(2\sum_m s_mm_{m,A}x^m\right)\,.
\label{eq:Omega0}
\end{gather}
This is called the master equation, and this determines $\Omega$ for a given $H_0$. Once $\Omega$ is obtained, we have $S$ up to phase.
But \(S\) can be fixed to be real by using the \(U(1)\) gauge symmetry, so that we find \(A_m=0\) from Eq. (\ref{eq:solofmodu}).
Then the real scalar field \(\varphi_m\) is given by
\begin{gather}
\varphi_m=\frac{1}{2}s_m\partial_m\log\Omega.
\label{eq:varphi}
\end{gather}
This is the solid procedure to get $\varphi_m$: give the moduli matrix $H_0$ and solve the master equation to get $\Omega$.
Note that there is a redundancy in Eq.~(\ref{eq:solofmodu}). The physical fields  \(H^A,A_m\) and \(\varphi_m\) for a given $(S,H_0)$ are unchanged under multiplication of an arbitrary non-zero complex constant $c$ as \((cS, cH_0^A)\).
Appropriately removing this redundancy, the resulting moduli space reads \( ( \mathbb{C}^{N_\mathrm{F}}-\{0\} )/\mathbb{C}^*\simeq \mathbb{C}P^{N_\mathrm{F}-1}\).

Before closing this subsection, we mention a constant shift of the mass vectors.
Consider a constant mass shift of all $\vec m_A$ as
\be
\vec m_A \to \vec m_A - \vec m_0\,,
\label{eq:mass_shift_boson}
\ee
with a constant vector \(\vec m_0 = (m_{1,0}, m_{2,0}, m_{3,0})\).
From Eqs. (\ref{eq:master}) and (\ref{eq:Omega0}) the solution \(\Omega\) of the master equation reads
\begin{gather}
\Omega \rightarrow \Omega
\exp\left(-2\sum_m s_m m_{m,0}x^m\right)\,.
\label{eq:omega_shift}
\end{gather}
Then the scalar field \(\varphi_m\) is shifted as
\begin{gather}
\varphi_m\rightarrow
\frac{1}{2}s_m\partial_m\log\left[ \Omega \exp\left(-2\sum_n s_n m_{0,n} x^n\right) \right] 
=\varphi_m - m_{0,m}\,.
\label{eq:phi_shift}
\end{gather}
This is, of course, expected from Eq.~(\ref{eq:Lagrangian_boson}) where any constant shift in $M_m$ can be absorbed by shifting $\varphi_m$.
Therefore, the constant shift yields no physical changes. Indeed, the constant shift does not affect the energy density of the solitons since ${\cal W}_m$ and ${\cal S}_{mn}$ consist only of derivatives of $\varphi_m$.
We will come back this point when we will include fermions in the later sections.

\subsection{Examples of the BPS monopole-string-domain wall composites}
\label{sec:example_MSD}

The master equation (\ref{eq:master}) cannot be solved analytically in general except for the special case \cite{Eto:2020vjm} where the model has the $S_4$ discrete symmetry. The other exception is the strong gauge coupling limit $e^2 \to \infty$ where the gauge field $A_\mu$ and the scalar field $\varphi_m$ are non-dynamical since they are infinitely massive (their kinetic terms vanish), and the model reduces to the massive non-linear sigma model whose target space is $\mathbb{C}P^{N_{\rm F}-1}$. There, the analytic solution for Eq.~(\ref{eq:master}) is $\Omega = \Omega_0$ \cite{Eto:2020vjm,Eto:2020cys}.
Namely, the analytic  solutions are only available for the special case or limits, and we have to numerically solve Eq.~(\ref{eq:master}) in general.
However, the topological properties of massless fermions do not depend on the details of the background solution, so the analytical solution at the limit $e^2 \to \infty$ is sufficient for the purposes of this study. Hence, we will take $\Omega = \Omega_0$ in the rest of this paper.
Note that we fix the signs as $s_m = +1$ ($m=1,2,3$) throughout this subsection. 
Solutions for the generic choice can be reproduced by changing the coordinate as $x^m \to s^m x^m$ together with $A_m \to s_m A_m$.

\begin{figure}[t]
\centering
\begin{minipage}[b]{0.24\hsize}
\centering
\includegraphics[width=0.9\textwidth]{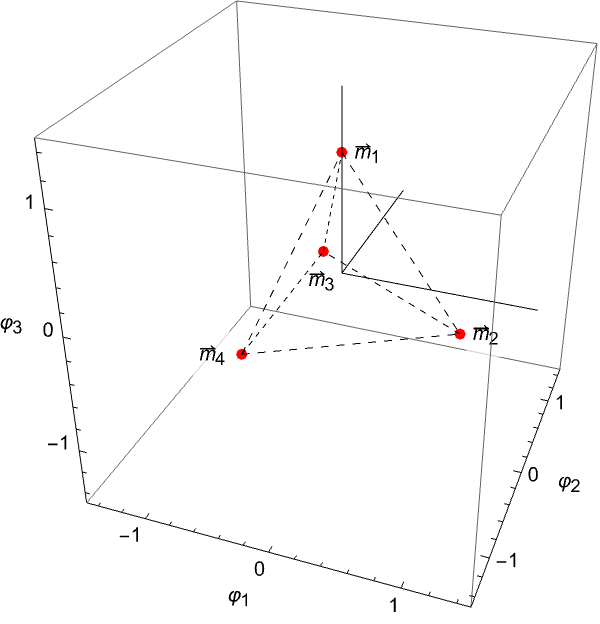}
\subcaption{}
\end{minipage}
\begin{minipage}[b]{0.24\hsize}
\centering
\includegraphics[width=0.9\textwidth]{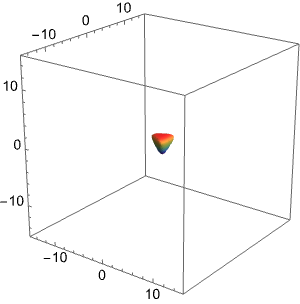}
\subcaption{}
\end{minipage}
\begin{minipage}[b]{0.24\hsize}
\centering
\includegraphics[width=0.9\textwidth]{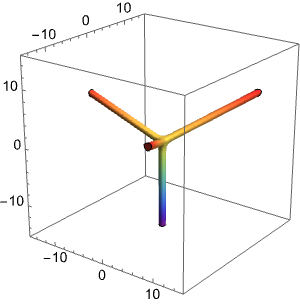}
\subcaption{}
\end{minipage}
\begin{minipage}[b]{0.24\hsize}
\centering
\includegraphics[width=0.9\textwidth]{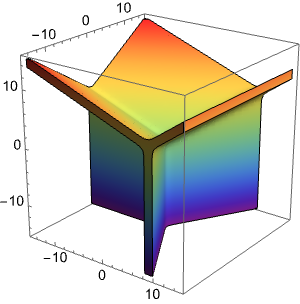}
\subcaption{}
\end{minipage}
\\
\centering
\begin{minipage}[b]{0.24\hsize}
\centering
\includegraphics[width=0.9\textwidth]{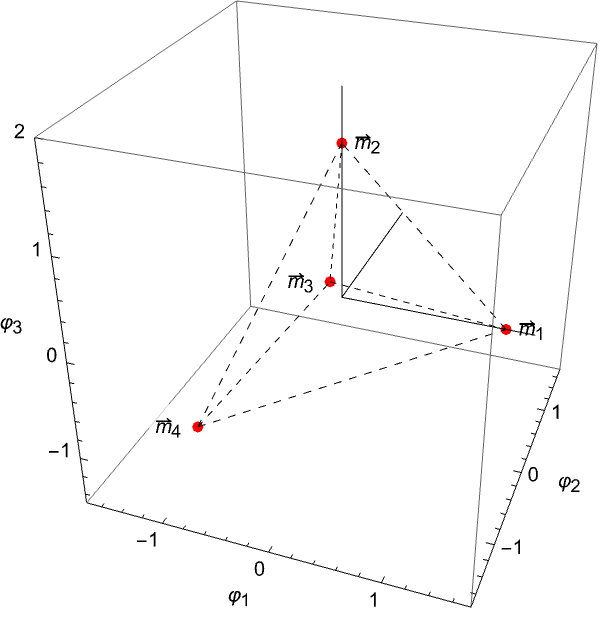}
\subcaption{}
\end{minipage}
\begin{minipage}[b]{0.24\hsize}
\centering
\includegraphics[width=0.9\textwidth]{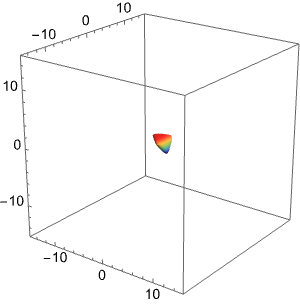}
\subcaption{}
\end{minipage}
\begin{minipage}[b]{0.24\hsize}
\centering
\includegraphics[width=0.9\textwidth]{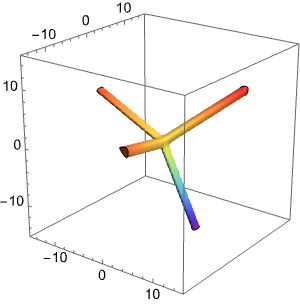}
\subcaption{}
\end{minipage}
\begin{minipage}[b]{0.24\hsize}
\centering
\includegraphics[width=0.9\textwidth]{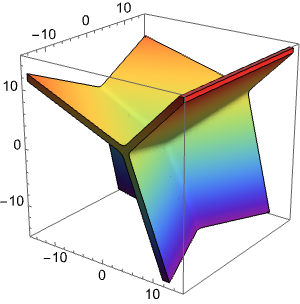}
\subcaption{}
\end{minipage}
\caption{
The BPS monopole-string-domain wall composites in the model with \(N_\mathrm{F}=4\).
Top panels are for (\ref{eq:massvectors1}) and bottom panels are for in (\ref{eq:massvectors2}).
(a) and (e) show the mass polyhedra,
(b) and (f) show monopole charge density \(\mathcal{M}\), (c) and (g) show the string charge density \(\mathcal{S}=S_{12}+S_{23}+S_{31}\), (d) and (h) show the domain wall charge density \(\mathcal{W}= {\cal W}_1 + {\cal W}_2 + {\cal W}_3\).}
\label{fig:3d_4vac_BG}
\end{figure}

Let us consider the case of \(N_\mathrm{F}=4\) which is minimum for a monopole-string-domain wall composite.
We consider two sets of the mass vectors: the one is a regular tetrahedron
\begin{gather}
\begin{split}
\vec{m}_{1}&=m\qty(0,0,1)\,, \\
\vec{m}_{2}&=m\qty(2\sqrt{2}/3, 0, -1/3)\,, \\
\vec{m}_{3}&=m\qty(-\sqrt{2}/3, \sqrt{2/3}, -1/3)\,, \\
\vec{m}_{4}&=m\qty(-\sqrt{2}/3, -\sqrt{2/3}, -1/3)\,,
\end{split}
\label{eq:massvectors1}
\end{gather}
and the other is a generic tetrahedron
\begin{gather}
\begin{split}
\vec{m}_{1}&=m\qty(3/2,0,0)\,, \\
\vec{m}_{2}&=m\qty(0,0,3/2)\,,\\
\vec{m}_{3}&=m\qty(-1/2,1,-1/2)\,, \\
\vec{m}_{4}&=m\qty(-1,-1,-1)\,,
\end{split}
\label{eq:massvectors2}
\end{gather}
with a parameter \(m\). Note that these are mere examples for exhibition purpose, and any tetrahedron is possible.
The four mass vectors are thought of as four vertices of the tetrahedron.
Now the model is set. Next, we fix the moduli matrix $H_0$ of the BPS solution. 
The desired monopole-string-domain wall composite is generated by the moduli matrix
$H_0=\qty(1,1,1,1)$.
We show the topological charge densities ${\cal M}$, ${\cal S}_{mn}$, and ${\cal W}_m$ 
in Fig.~\ref{fig:3d_4vac_BG}.  The first row corresponds to the mass vector in Eq.~(\ref{eq:massvectors1}), and 
the second row to those in Eq.~(\ref{eq:massvectors2}). The former has the tetrahedral symmetry while the latter does not. But they are qualitatively same: they consist of one monopole, four strings, six domain walls, and four vacua. This structure clearly reflects the mass tetrahedron which has one body, four faces, six edges, and four vertices.

\begin{figure}[t]
\centering
\begin{minipage}[b]{0.24\hsize}
\centering
\includegraphics[width=0.9\textwidth]{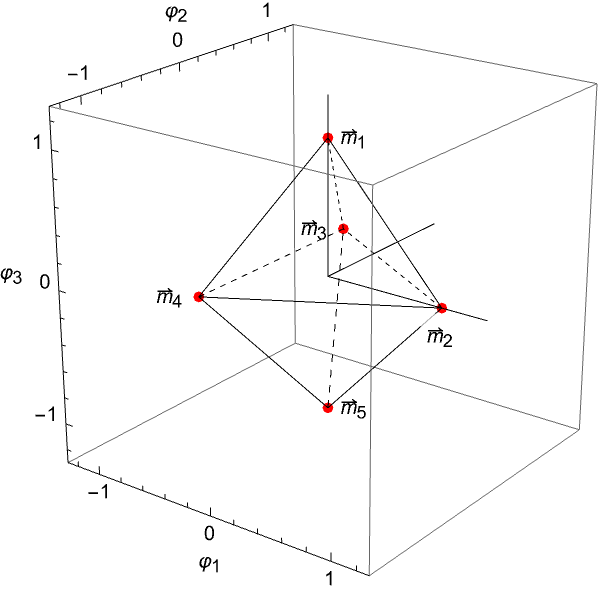}
\subcaption{}
\end{minipage}
\begin{minipage}[b]{0.24\hsize}
\centering
\includegraphics[width=0.9\textwidth]{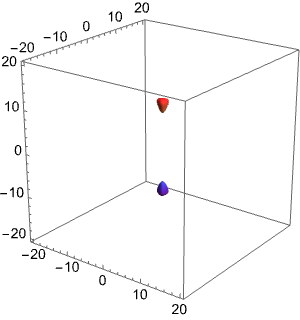}
\subcaption{}
\end{minipage}
\begin{minipage}[b]{0.24\hsize}
\centering
\includegraphics[width=0.9\textwidth]{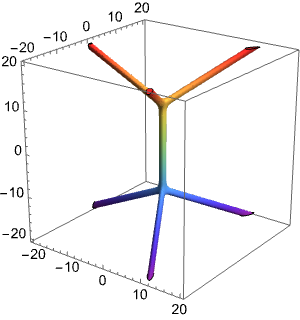}
\subcaption{}
\end{minipage}
\begin{minipage}[b]{0.24\hsize}
\centering
\includegraphics[width=0.9\textwidth]{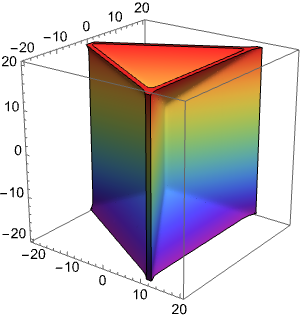}
\subcaption{}
\end{minipage}
\\
\centering
\begin{minipage}[b]{0.24\hsize}
\centering
\includegraphics[width=0.9\textwidth]{3D_susy_5vac_massvector_branch.png}
\subcaption{}
\end{minipage}
\begin{minipage}[b]{0.24\hsize}
\centering
\includegraphics[width=0.9\textwidth]{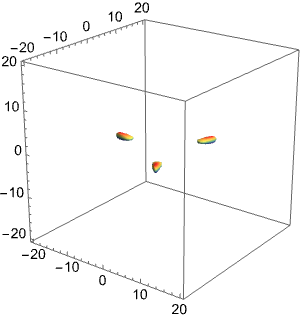}
\subcaption{}
\end{minipage}
\begin{minipage}[b]{0.24\hsize}
\centering
\includegraphics[width=0.9\textwidth]{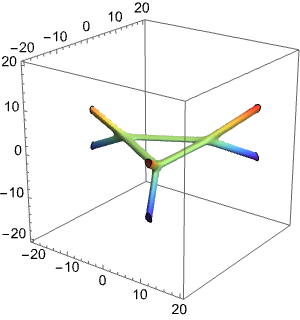}
\subcaption{}
\end{minipage}
\begin{minipage}[b]{0.24\hsize}
\centering
\includegraphics[width=0.9\textwidth]{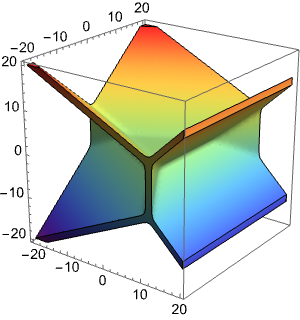}
\subcaption{}
\end{minipage}
\\
\centering
\begin{minipage}[b]{0.24\hsize}
\centering
\includegraphics[width=0.9\textwidth]{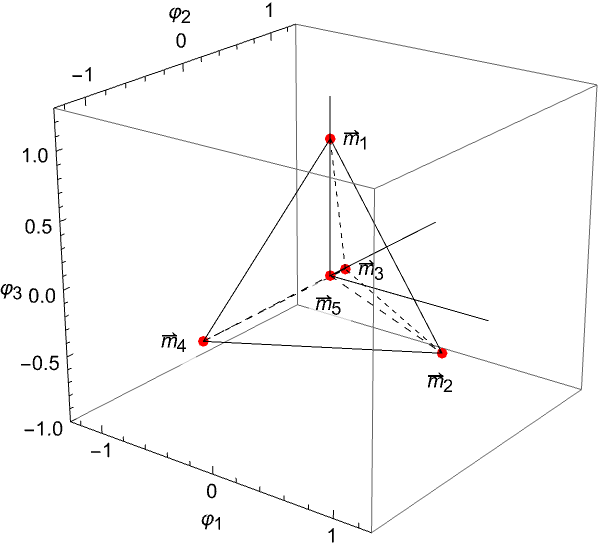}
\subcaption{}
\end{minipage}
\begin{minipage}[b]{0.24\hsize}
\centering
\includegraphics[width=0.9\textwidth]{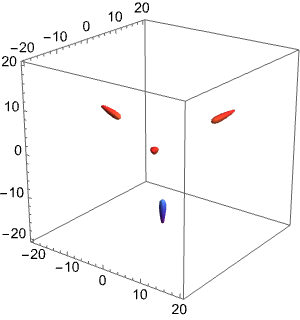}
\subcaption{}
\end{minipage}
\begin{minipage}[b]{0.24\hsize}
\centering
\includegraphics[width=0.9\textwidth]{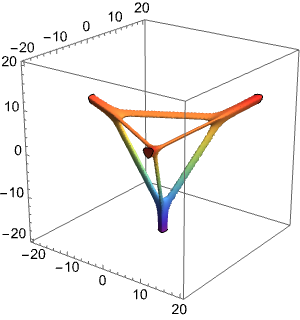}
\subcaption{}
\end{minipage}
\begin{minipage}[b]{0.24\hsize}
\centering
\includegraphics[width=0.9\textwidth]{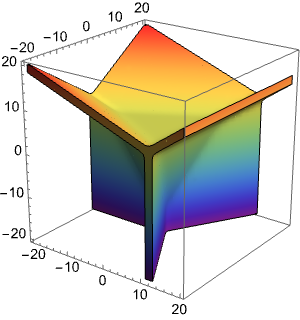}
\subcaption{}
\end{minipage}
\caption{
The BPS monopole-string-domain wall composites in the model with \(N_\mathrm{F}=5\).
The panels on the first and second rows are for the mass vectors arranged on the vertices of a hexahedron in Eq.~(\ref{eq:mass_hexahedron}), 
and those of the bottom row are for those placed on the vertices of a tetrahedron in Eq.~(\ref{eq:massvectors1}) in addition to the origin.
The left-most column shows the mass polyhedra, the second column shows ${\cal M}$, the third column shows 
\(\mathcal{S}=S_{12}+S_{23}+S_{31}\), and the right-most column shows \(\mathcal{W}= {\cal W}_1 + {\cal W}_2 + {\cal W}_3\).
}
\label{fig:3d_5vac_BG}
\end{figure}

More complicated configurations can be constructed in \(N_\mathrm{F}\geq5\) case.
Here we show two simplest examples for \(N_\mathrm{F}=5\).
The mass vectors for the first example are arranged on  vertices of a hexahedron
\begin{gather}
\begin{split}
\vec{m}_{1}&=m\qty(0,0,4/3)\,, \\
\vec{m}_{2}&=m\qty(2\sqrt{2}/3, 0, 0)\,, \\
\vec{m}_{3}&=m\qty(-\sqrt{2}/3, \sqrt{2/3}, 0)\,, \\
\vec{m}_{4}&=m\qty(-\sqrt{2}/3, -\sqrt{2/3}, 0)\,,\\
\vec{m}_{5}&=m\qty(0,0,-4/3)\,.
\end{split}
\label{eq:mass_hexahedron}
\end{gather}
An analytic solution with \(H_0=(1,\expon^{10},\expon^{10},\expon^{10},1)\) is shown in the first row of Fig.~\ref{fig:3d_5vac_BG}.
The solution consists of two monopoles, six semi-infinite strings and one finite string, nine domain walls, and five vacua.
With the same mass vectors, but changing the moduli matrix as \(H_0=(\expon^5,1,1,1,\expon^5)\) gives us a different solution shown in the second
row of Fig.~\ref{fig:3d_5vac_BG}. The solution has three monopoles, three finite and six semi-infinite strings, and 
nine semi-infinite and 
one finite domain walls.
The third example has the four mass vectors given in Eq.~(\ref{eq:massvectors1}) corresponding to the regular tetrahedron accompanied with
the fifth vector $\vec m_5 = \vec 0$. A solution for \(H_0=(1,1,1,1,\expon^5)\) is shown in the third row of Fig.~\ref{fig:3d_5vac_BG}. The composite soliton has four monopoles, three semi-infinite and six finite strings, ten domain walls, and five vacua.

\begin{figure}[h]
\centering
\begin{minipage}[b]{0.3\hsize}
\centering
\includegraphics[width=0.9\textwidth]{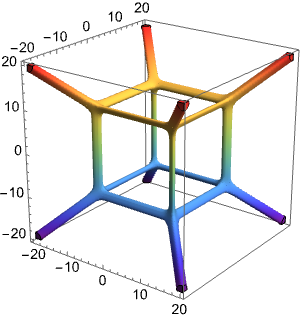}
\end{minipage}
\begin{minipage}[b]{0.3\hsize}
\centering
\includegraphics[width=0.9\textwidth]{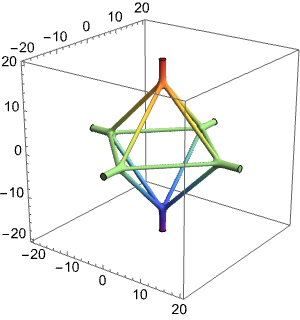}
\end{minipage}
\caption{
Composite solitons for \(N_\mathrm{F}=7\) and \(N_\mathrm{F}=9\), respectively.
Only strings \(\mathcal{S}=S_{12}+S_{23}+S_{31}\) are shown.
}
\label{fig:3d_polyhedron_BG}
\end{figure}

More complicated network solutions can be easily constructed as shown in Fig. \ref{fig:3d_polyhedron_BG}.

\section{Monopole-string-domain wall fermions and polyhedral vacuum fermions}
\label{sec:MSDWF}

\subsection{Fermion zero mode functions}
\label{sec:fermion_zero_mode}

Let us next investigate fermion zero modes in the background of three dimensional 
monopole-string-domain wall composites reviewed in the previous section.
The fermion zero modes in the background of two dimensional string-domain wall composites 
were recently investigated by the same authors in \cite{Eto:2023orr}. 
We would like to clarify how previous results on two dimensional string-domain wall fermions change 
if we replace the background soliton with a three dimensional monopole-string-domain wall.

We introduce Dirac fermions $\Psi_a$ ($a=1,2$) which couples with the real scalar fields 
$\varphi_m$ ($m=1,2,3$) through a Yukawa term as
\begin{gather}
\mathcal{L}_\mathrm{F}=i\bar{\Psi}_a\gamma^\mu\partial_\mu\Psi_a-h\bar{\Psi}_a \Phi_{ab} \Psi_b,\quad
\Phi_{ab} = \sum_m T^m_{ab}\varphi_m\,,
\label{eq:fermionLagrangian}
\end{gather}
where \(h\) is a Yukawa coupling constant and we assume $h>0$.
Here, \(T_{ab}^m=\frac{1}{2}(\sigma^m)_{ab}\) is a generator of $SU(2)$.  
The fermionic Lagrangian is invariant under the $SU(2)$ transformation,  
$\Psi \to G \Psi$ and $\Phi \to G \Phi G^\dag$ with $G \in SU(2)$.
Note, however, that the $SU(2)$ symmetry is explicitly broken in the bosonic Lagrangian 
as we mentioned at the end of Sec.~\ref{sec:sec:GMSW_1}.
The gamma matrices are given by
\begin{gather}
\gamma^0=
\begin{pmatrix}
0 & -i \\
i & 0
\end{pmatrix}
,\quad\gamma^m=
\begin{pmatrix}
-i\sigma^m & 0 \\
0 & i\sigma^m
\end{pmatrix}
.
\end{gather}

The Dirac equation is given by
\begin{gather}
i\gamma^\mu\partial_\mu\Psi_a-\frac{h}{2}\sigma_{ab}^m\varphi_m\Psi_b=0.
\label{eq:Diraceq}
\end{gather}
Let us separate $x^0$ and $x^m$  as 
\begin{gather}
\begin{pmatrix}
\chi_a(x^\mu) \\
\bar{\xi}_a(x^\mu)
\end{pmatrix}
=\expon^{-i\omega t}
\begin{pmatrix}
\chi_{a}^{(\omega)}(x^m) \\
\bar{\xi}_a^{(\omega)}(x^m)
\end{pmatrix}
,
\end{gather}
where $\chi_a$ and $\bar \xi_a$ are the two component Weyl spinors.
Then the Dirac equation can be expressed as the following 8 by 8 matrix form
\begin{gather}
\left(
\begin{array}{cccc} 
\sigma^m\partial_m-\frac{h}{2}\varphi_3 & 0  & -\frac{h}{2}(\varphi_1-i\varphi_2) & 0 \\
0 & -\sigma^m\partial_m-\frac{h}{2}\varphi_3 & 0 & -\frac{h}{2}(\varphi_1-i\varphi_2) \\
-\frac{h}{2}(\varphi_1+i\varphi_2) & 0 & \sigma^m\partial_m+\frac{h}{2}\varphi_3 & 0 \\
0 & -\frac{h}{2}(\varphi_1+i\varphi_2) & 0 & -\sigma^m\partial_m+\frac{h}{2}\varphi_3
\end{array}
 \right)
\left(
\begin{array}{c}
\chi_1^{(\omega)} \\
\bar{\xi}_1^{(\omega)} \\
\chi_2^{(\omega)} \\
\bar{\xi}_2^{(\omega)}
\end{array}
\right)
=i\omega
\begin{pmatrix}
\bar{\xi}_1^{(\omega)} \\
-\chi_1^{(\omega)} \\
\bar{\xi}_2^{(\omega)} \\
-\chi_2^{(\omega)} \\
\end{pmatrix}
.
\label{eq:Diraceq_matrix}
\end{gather}
Since we are interested in a zero mode, we will set \(\omega=0\) in what follows.
Then the above equation can be further decomposed into two 4 by 4 matrices as
\begin{align}
\begin{pmatrix}
\sigma^m\partial_m-\frac{h}{2}\varphi_3 & -\frac{h}{2}(\varphi_1-i\varphi_2) \\
-\frac{h}{2}(\varphi_1+i\varphi_2) & \sigma^m\partial_m+\frac{h}{2}\varphi_3
\end{pmatrix}
\begin{pmatrix}
\chi_1^{(0)} \\
\chi_2^{(0)}
\end{pmatrix}
&=0, \label{eq:Diraceq1} \\
\begin{pmatrix}
-\sigma^m\partial_m-\frac{h}{2}\varphi_3 & -\frac{h}{2}(\varphi_1-i\varphi_2) \\
-\frac{h}{2}(\varphi_1+i\varphi_2) & -\sigma^m\partial_m+\frac{h}{2}\varphi_3
\end{pmatrix}
\begin{pmatrix}
\bar{\xi}_1^{(0)} \\
\bar{\xi}_2^{(0)}
\end{pmatrix}
&=0.
\label{eq:Diraceq2}
\end{align}

We now solve these equations with the bosonic background solution $\varphi_m$ given in Eq.~(\ref{eq:varphi}).
In order to solve these equations, let us make an ansatz for \(\chi_a\) and \(\bar{\xi}_a\)  as\footnote{
A massless fermion for a spherically symmetric global monopole is well known \cite{Jackiw:1975fn} and it is reviewed in the appendix \ref{sec:appA}.
}
\begin{gather}
\chi_{a\alpha}^{(0)}(x^m)=f(x^m)\epsilon_{a\alpha}
,\quad \bar{\xi}_a^{\dot{\alpha}(0)}(x^m)=g(x^m)\epsilon^{a\dot{\alpha}},
\label{eq:ansatz_fermion}
\end{gather}
where $\alpha$ and $\dot \alpha$ are the Weyl spinor indices, 
and $a$ is $SU(2)$ isospinor index with the antisymmetric tensor $\epsilon_{12} = \epsilon^{12} = 1$.
Plugging Eq.~\refer{eq:ansatz_fermion} into Eqs.~(\ref{eq:Diraceq1}) and (\ref{eq:Diraceq2}), we have the following equations
\begin{align}
\begin{pmatrix}
\partial_3-\frac{h}{2}\varphi_3 & \partial_1-i\partial_2 & -\frac{h}{2}(\varphi_1-i\varphi_2) & 0 \\
\partial_1+i\partial_2 & -\partial_3-\frac{h}{2}\varphi_3 & 0 & -\frac{h}{2}(\varphi_1-i\varphi_2) \\
-\frac{h}{2}(\varphi_1+i\varphi_2) & 0 & \partial_3+\frac{h}{2}\varphi_3 & \partial_1-i\partial_2 \\
0 & -\frac{h}{2}(\varphi_1+i\varphi_2) & \partial_1+i\partial_2 & -\partial_3+\frac{h}{2}\varphi_3
\end{pmatrix}
\begin{pmatrix}
0 \\
f \\
-f \\
0
\end{pmatrix}
&=0, \label{eq:Diraceq_f} \\
\begin{pmatrix}
-\partial_3-\frac{h}{2}\varphi_3 & -\partial_1+i\partial_2 & -\frac{h}{2}(\varphi_1-i\varphi_2) & 0 \\
-\partial_1-i\partial_2 & \partial_3-\frac{h}{2}\varphi_3 & 0 & -\frac{h}{2}(\varphi_1-i\varphi_2) \\
-\frac{h}{2}(\varphi_1+i\varphi_2) & 0 & -\partial_3+\frac{h}{2}\varphi_3 & -\partial_1+i\partial_2 \\
0 & -\frac{h}{2}(\varphi_1+i\varphi_2) & -\partial_1-i\partial_2 & \partial_3+\frac{h}{2}\varphi_3
\end{pmatrix}
\begin{pmatrix}
0 \\
g \\
-g \\
0
\end{pmatrix}
&=0,
\label{eq:Diraceq_g}
\end{align}
where the elements of the column vectors are placed as  \((\chi_1,\chi_2)^T=(\chi_{11},\chi_{12},\chi_{21},\chi_{22})^T$, 
and $\left(\bar{\xi}_1,\bar{\xi}_2\right)^T = \left(\bar{\xi}_1^{\dot{1}},\bar{\xi}_1^{\dot{2}},\bar{\xi}_2^{\dot{1}},\bar{\xi}_2^{\dot{2}}\right)^T\).
Fortunately, these are expressed in the following compact equations
\begin{gather}
\qty(\partial_m+\frac{h}{2}\varphi_m)f=0,\quad\qty(\partial_m-\frac{h}{2}\varphi_m)g=0.
\label{eq:Diraceq_reduced}
\end{gather}
Suppose \(f\) and \(g\) are both non zero, then $\varphi_m$ is written as
\begin{gather}
\varphi_m=\partial_m\qty(-\frac{2}{h}\log f)=\partial_m\qty(\frac{2}{h}\log g).
\label{eq:phi_by_fg}
\end{gather}
Therefore, we have
\begin{gather}
g \propto f^{-1}\,.
\end{gather}
This means that if $f$ is normalizable then $g$ is not, and vice versa.
Therefore, either \(f\) or \(g\) must be zero  in Eq.~(\ref{eq:ansatz_fermion}).
To proceed, recall $\varphi_m$ in the expression in Eq.~(\ref{eq:varphi}), and compare it with Eq.~(\ref{eq:phi_by_fg}).
Then we find the relations between \(f,g\) and \(\Omega\)
\begin{gather}
\frac{1}{2}s_m\log\Omega=-\frac{2}{h}\log f + C\,,\qquad g = 0\,,\quad
\end{gather}
or
\begin{gather}
\frac{1}{2}s_m\log\Omega=\frac{2}{h}\log g + C\,,\qquad f= 0\,,
 \end{gather}
with $C$ being a constant.
In both cases, the left hand side has the index $m$, but the right hand side does not. 
Therefore, only possible solutions can have $\vec s = (1,1,1)$ or $(-1,-1,-1)$.
Taking into account the fact that $\Omega$ is real positive and linearly diverge in $x^m$, 
we conclude that there is one zero mode function that can be normalized as
\begin{gather}
\vec s = (1,1,1):\quad f = \Omega^{-\frac{h}{4}},\quad g = 0\,,\quad
\Psi_1 =
\begin{pmatrix}
0 \\
\Omega^{-\frac{h}{4}} \\
0 \\
0
\end{pmatrix}
,\quad
\Psi_2 =
\begin{pmatrix}
-\Omega^{-\frac{h}{4}} \\
0 \\
0 \\
0
\end{pmatrix}\,,
\label{eq:+++}
\end{gather}
and 
\begin{gather}
\vec s = (-1,-1,-1):\quad g = \Omega^{-\frac{h}{4}},\quad f = 0\,,
\quad
\Psi_1 =
\begin{pmatrix}
0 \\
0 \\
0 \\
\Omega^{-\frac{h}{4}}
\end{pmatrix}
,\quad \Psi_2 =
\begin{pmatrix}
0 \\
0 \\
-\Omega^{-\frac{h}{4}} \\
0
\end{pmatrix}\,.
\end{gather}

We could not find any zero modes for $\vec s \neq \pm (1,1,1)$.
It does not mean non-existence of fermion zero modes for the background solutions for $\vec s \neq \pm (1,1,1)$, but just the ansatz (\ref{eq:ansatz_fermion}) is not appropriate.
Let us make another ansatz as
\begin{gather}
\chi_{a\alpha}^{(0)}=
\begin{pmatrix}
0 \\
f \\
f \\
0
\end{pmatrix}
,\quad \bar{\xi}_a^{\dot{\alpha}(0)}=0,\quad \mathrm{or}\quad \chi_{a\alpha}^{(0)}=0,\quad \bar{\xi}_a^{\dot{\alpha}(0)}=
\begin{pmatrix}
0 \\
g \\
g \\
0
\end{pmatrix}
. \label{eq:ansatz_fermion2}
\end{gather}
Plugging the first ansatz into Eq. (\ref{eq:Diraceq1}) and (\ref{eq:Diraceq2}), the Dirac equation reduces to
\be
\p_m f + \tau_m \frac{h}{2}\varphi_m f = 0
\quad \Leftrightarrow \quad
\varphi_m = - \tau_m \frac{2}{h} \p_m \log f\,,
\label{eq:tau_1}
\ee
with $\vec \tau = (-1,-1,1)$.
Similarly, if we substitute the second one into Eq. (\ref{eq:Diraceq1}) and (\ref{eq:Diraceq2}), we end up with
\be
\p_m g - \tau_m \frac{h}{2}\varphi_m g = 0
\quad \Leftrightarrow \quad
\varphi_m = \tau_m \frac{2}{h} \p_m \log g\,.
\label{eq:tau_2}
\ee
These should be again compared with $\varphi_m = s_m\frac{1}{2}\p_m \log \Omega$ in Eq.~(\ref{eq:varphi}), 
and consistency on the index structure allows us to choose $s_m = \pm \tau_m$. The choice of $\pm$ is fixed by normalizability condition,
resulting in $\vec s = \vec\tau$ for Eq.~(\ref{eq:tau_1}) and  $\vec s = - \vec\tau$ for Eq.~(\ref{eq:tau_2}).
In summary, we find the fermion zero mode function for another $\vec s$ as
\begin{alignat}{3}
&(s_1,s_2,s_3)=(-1,-1,1)\,,\quad
&\Psi_1=
\begin{pmatrix}
0 \\
\Omega^{-\frac{h}{4}} \\
0 \\
0
\end{pmatrix}
,\quad
&\Psi_2 =
\begin{pmatrix}
\Omega^{-\frac{h}{4}} \\
0 \\
0 \\
0
\end{pmatrix}\,, \\
&(s_1,s_2,s_3)=(1,1,-1)\,,\quad
&\Psi_1 =
\begin{pmatrix}
0 \\
0 \\
0 \\
\Omega^{-\frac{h}{4}}
\end{pmatrix}
,\quad
&\Psi_2 =
\begin{pmatrix}
0 \\
0 \\
\Omega^{-\frac{h}{4}} \\
0
\end{pmatrix}\,.
\label{eq:++-}
\end{alignat}
The fermion zero modes for the remaining possible combinations of $\vec s$ can be easily obtained in a similar manner.
Below we only show the results.
\begin{alignat}{3}
&(s_1,s_2,s_3)=(1,-1,-1)\,,\quad
&\Psi_1=
\begin{pmatrix}
\Omega^{-\frac{h}{4}} \\
0 \\
0 \\
0
\end{pmatrix}
,\quad
&\Psi_2 =
\begin{pmatrix}
0\\
- \Omega^{-\frac{h}{4}} \\
0 \\
0
\end{pmatrix}\,, \\
&(s_1,s_2,s_3)=(-1,1,1)\,,\quad
&\Psi_1 =
\begin{pmatrix}
0 \\
0 \\
\Omega^{-\frac{h}{4}} \\
0
\end{pmatrix}
,\quad
&\Psi_2 =
\begin{pmatrix}
0 \\
0 \\
0 \\
-\Omega^{-\frac{h}{4}}
\end{pmatrix}\,, \\
&(s_1,s_2,s_3)=(-1,1,-1)\,,\quad
&\Psi_1=
\begin{pmatrix}
\Omega^{-\frac{h}{4}} \\
0 \\
0 \\
0
\end{pmatrix}
,\quad
&\Psi_2 =
\begin{pmatrix}
0\\
\Omega^{-\frac{h}{4}} \\
0 \\
0
\end{pmatrix}\,, \\
&(s_1,s_2,s_3)=(1,-1,1)\,,\quad
&\Psi_1 =
\begin{pmatrix}
0 \\
0 \\
\Omega^{-\frac{h}{4}} \\
0
\end{pmatrix}
,\quad
&\Psi_2 =
\begin{pmatrix}
0 \\
0 \\
0 \\
\Omega^{-\frac{h}{4}}
\end{pmatrix}\,.
\end{alignat}

Let us summarize what we achieved in this subsection.
We succeeded in obtaining the fermion  zero mode function for all the BPS monopole-string-domain wall background with all possible choice of $\vec s$.
The zero mode function universally takes the form $\Omega^{-h/4}$. We emphasize that our solution is always valid regardless of whether the background solution $\Omega$ is obtained analytically or numerically. In any case, once $\Omega$ is determined by solving the master equation (\ref{eq:master}), we can obtain the fermion zero mode function $\Omega^{-h/4}$ directly without further calculations.

\subsection{Examples}

Let us show several concrete fermion zero mode functions. 
As is explained in the previous subsection, the zero mode function is $\Omega^{-h/4}$ regardless of choice of $\vec s$.
So we only show $\Omega^{-h/4}$ below for the concrete solutions that we have obtained in Sec.~\ref{sec:example_MSD}.

\begin{figure}[htbp]
\centering
\begin{minipage}[b]{0.4\hsize}
\centering
\includegraphics[width=0.75\textwidth]{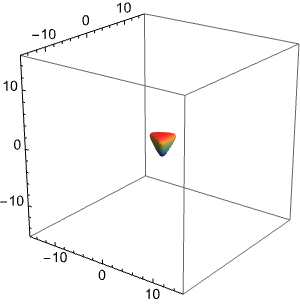}
\subcaption{The regular tetrahedron case}
\end{minipage}
\begin{minipage}[b]{0.4\hsize}
\centering
\includegraphics[width=0.75\textwidth]{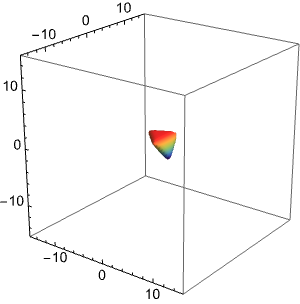}
\subcaption{The modified tetrahedron case}
\end{minipage}
\caption{The fermion zero mode functions \(\Omega^{-h/4}\) coupled with the monopole-string-domain wall composites given in Fig. \ref{fig:3d_4vac_BG}.
}
\label{fig:3d_4vac_fermion}
\end{figure}

The first example is the monopole-string-domain wall composites shown in Fig.~\ref{fig:3d_4vac_BG} for $N_{\rm F} = 4$.
It includes two solutions: the one has regular tetrahedral symmetry (the first row of Fig.~\ref{fig:3d_4vac_BG}) 
whereas the other (the second row of Fig.~\ref{fig:3d_4vac_BG}) does not.
Corresponding fermion zero mode functions are shown in Fig.~\ref{fig:3d_4vac_fermion}.
We observe that they are localized around the monopoles, and the zero mode functions have almost 
same shape as the monopole charge densities.

\begin{figure}[htbp]
\centering
\begin{minipage}[b]{0.32\hsize}
\centering
\includegraphics[width=0.75\textwidth]{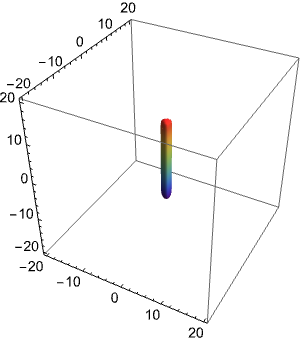}
\subcaption{}
\end{minipage}
\begin{minipage}[b]{0.32\hsize}
\centering
\includegraphics[width=0.75\textwidth]{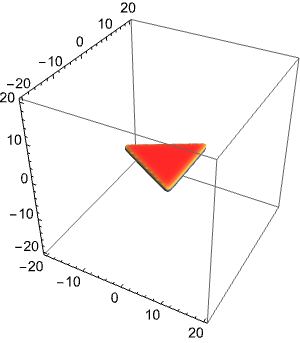}
\subcaption{}
\end{minipage}
\begin{minipage}[b]{0.32\hsize}
\centering
\includegraphics[width=0.75\textwidth]{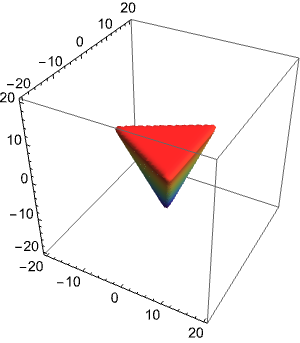}
\subcaption{}
\end{minipage}
\caption{The panel (a) [(b), (c)] shows the fermion zero mode function \(\Omega^{-h/4}\) for the first (second, third) row of Fig.~\ref{fig:3d_5vac_BG}.}
\label{fig:3d_5vac_fermion}
\end{figure}

The second example is the background solitons in the case of \(N_\mathrm{F}=5\) shown in Fig.~\ref{fig:3d_5vac_BG}.
We considered three different types given in Fig.~\ref{fig:3d_5vac_BG}.
The fermion zero mode for the first row of Fig.~\ref{fig:3d_5vac_BG} is not localized on either of the two monopoles, 
but on the inner string connecting the two monopoles, see Fig.~\ref{fig:3d_5vac_fermion}(a).
It for the second row of Fig.~\ref{fig:3d_5vac_BG} is not localized on either of the monopoles or strings but on the inner triangle domain wall
as shown in Fig.~\ref{fig:3d_5vac_fermion}(b).
The zero mode for the third row of Fig.~\ref{fig:3d_5vac_BG}  is not localized on either of any solitons, monopoles, strings, and domain walls, but throughout the central internal vacuum region as shown in Fig.~\ref{fig:3d_5vac_fermion}(c).

\begin{figure}[htbp]
\centering
\begin{minipage}[b]{0.4\hsize}
\centering
\includegraphics[width=0.75\textwidth]{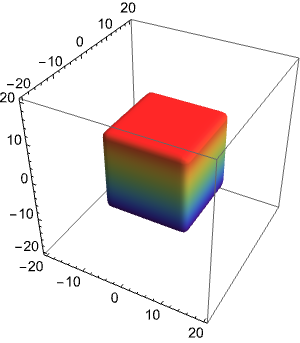}
\subcaption{}
\end{minipage}
\begin{minipage}[b]{0.4\hsize}
\centering
\includegraphics[width=0.75\textwidth]{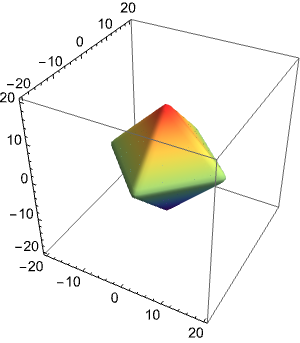}
\subcaption{}
\end{minipage}
\caption{
The panel (a) [(b)] shows the fermion zero mode function \(\Omega^{-h/4}\) for the background solitons of Fig.~\ref{fig:3d_polyhedron_BG}.}
\label{fig:3d_polyhedron_fermion}
\end{figure}

Finally, we show in Fig.~\ref{fig:3d_polyhedron_fermion} 
the fermion zero mode functions for the background solutions in Fig. \ref{fig:3d_polyhedron_BG}.
Similarly to Fig.~\ref{fig:3d_5vac_fermion}(c), the mode functions are localized throughout the central internal vacuum regions whose shapes are
a cube and an octahedron, respectively.

\begin{figure}[htbp]
\centering
\begin{minipage}[b]{0.16\hsize}
\centering
\includegraphics[width=0.99\textwidth]{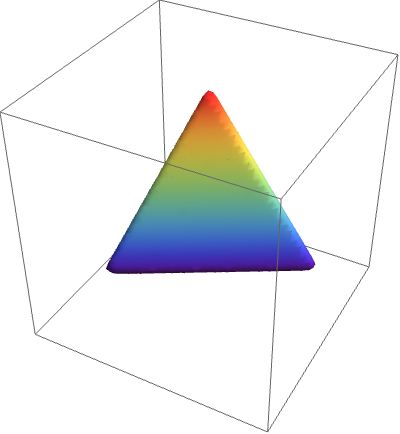}
\end{minipage}
\begin{minipage}[b]{0.16\hsize}
\centering
\includegraphics[width=0.99\textwidth]{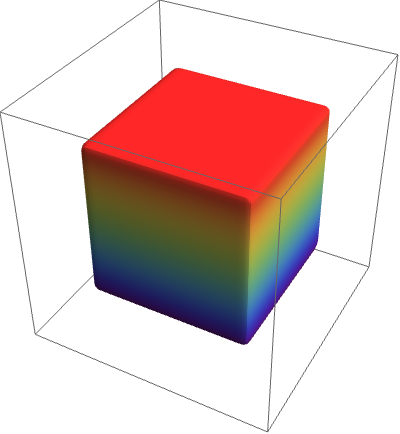}
\end{minipage}
\begin{minipage}[b]{0.16\hsize}
\centering
\includegraphics[width=0.99\textwidth]{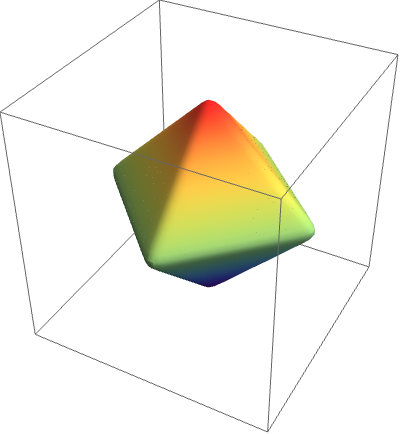}
\end{minipage}
\begin{minipage}[b]{0.16\hsize}
\centering
\includegraphics[width=0.99\textwidth]{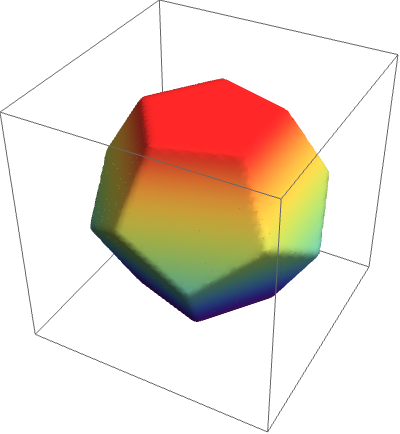}
\end{minipage}
\begin{minipage}[b]{0.16\hsize}
\centering
\includegraphics[width=0.99\textwidth]{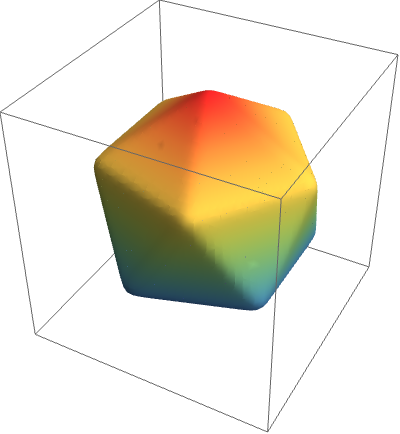}
\end{minipage}
\centering
\begin{minipage}[b]{0.16\hsize}
\centering
\includegraphics[width=0.99\textwidth]{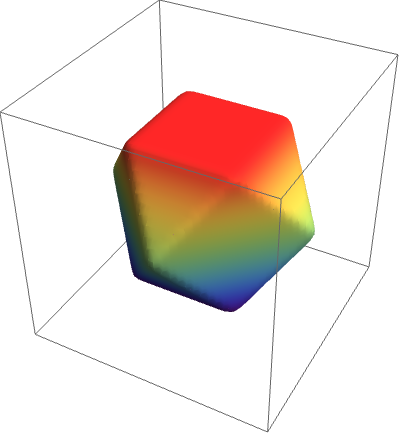}
\end{minipage}
\begin{minipage}[b]{0.16\hsize}
\centering
\includegraphics[width=0.99\textwidth]{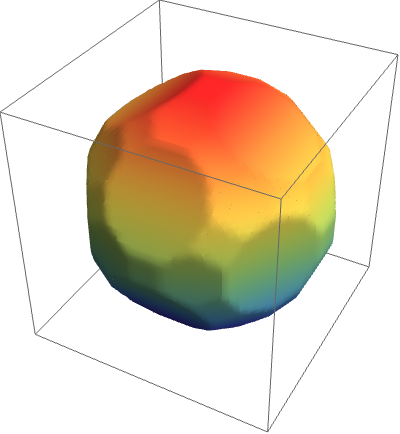}
\end{minipage}
\begin{minipage}[b]{0.16\hsize}
\centering
\includegraphics[width=0.99\textwidth]{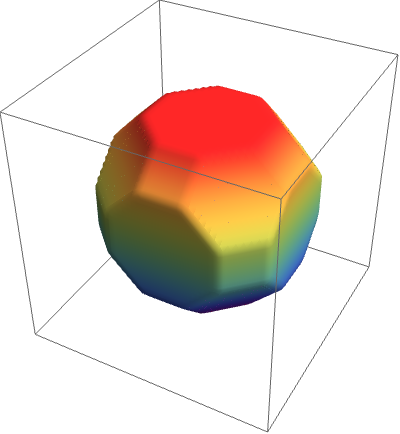}
\end{minipage}
\begin{minipage}[b]{0.16\hsize}
\centering
\includegraphics[width=0.99\textwidth]{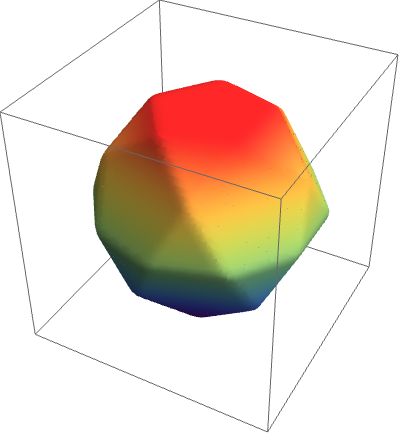}
\end{minipage}
\begin{minipage}[b]{0.16\hsize}
\centering
\includegraphics[width=0.99\textwidth]{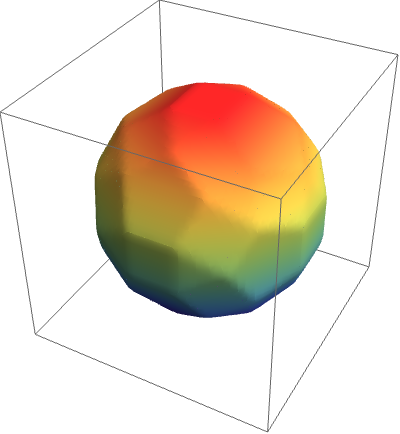}
\end{minipage}
\centering
\begin{minipage}[b]{0.16\hsize}
\centering
\includegraphics[width=0.99\textwidth]{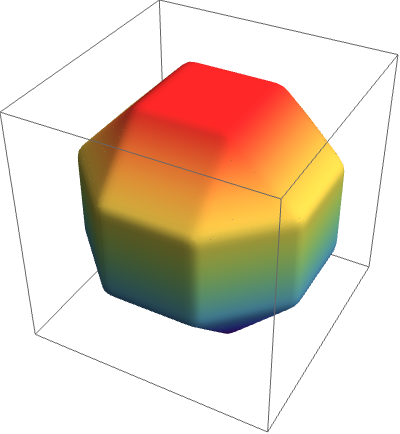}
\end{minipage}
\begin{minipage}[b]{0.16\hsize}
\centering
\includegraphics[width=0.99\textwidth]{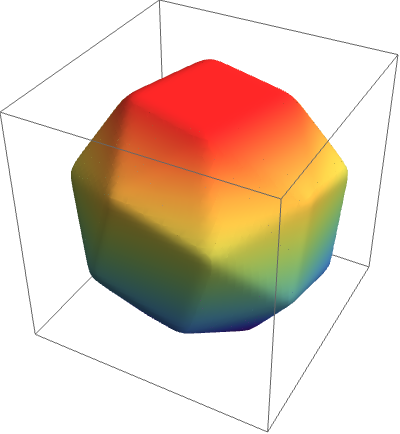}
\end{minipage}
\begin{minipage}[b]{0.16\hsize}
\centering
\includegraphics[width=0.99\textwidth]{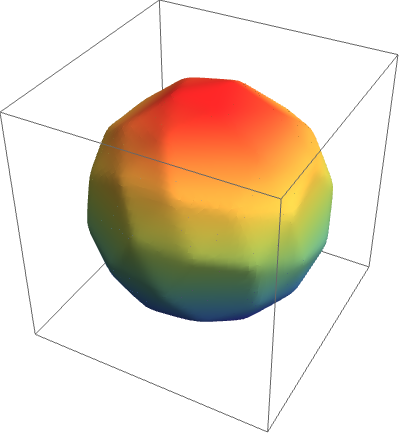}
\end{minipage}
\begin{minipage}[b]{0.16\hsize}
\centering
\includegraphics[width=0.99\textwidth]{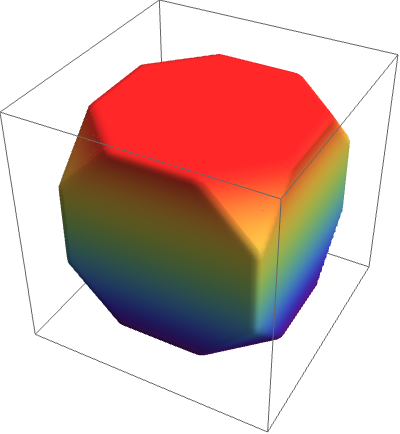}
\end{minipage}
\begin{minipage}[b]{0.16\hsize}
\centering
\includegraphics[width=0.99\textwidth]{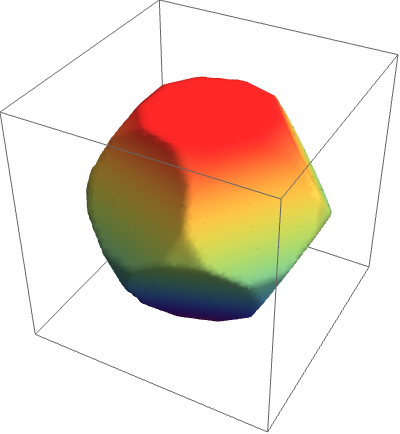}
\end{minipage}
\centering
\begin{minipage}[b]{0.16\hsize}
\centering
\includegraphics[width=0.99\textwidth]{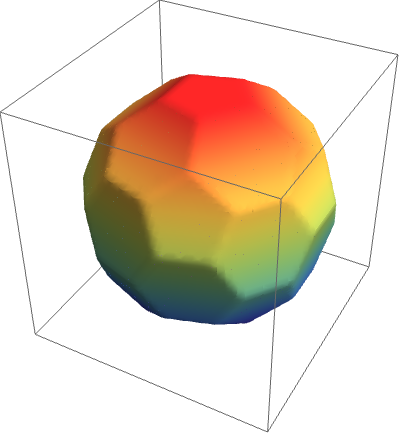}
\end{minipage}
\begin{minipage}[b]{0.16\hsize}
\centering
\includegraphics[width=0.99\textwidth]{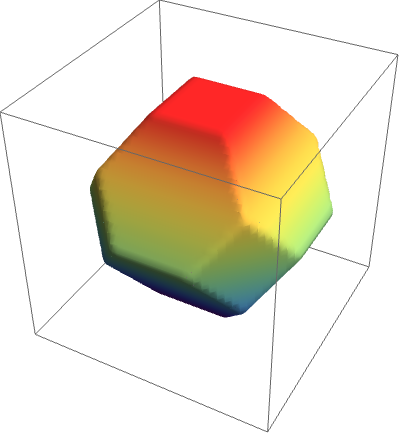}
\end{minipage}
\begin{minipage}[b]{0.16\hsize}
\centering
\includegraphics[width=0.99\textwidth]{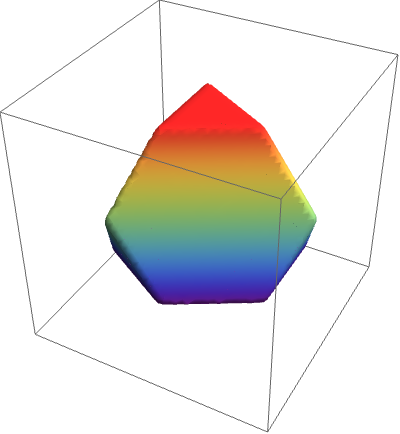}
\end{minipage}
\begin{minipage}[b]{0.16\hsize}
\centering
\includegraphics[width=0.99\textwidth]{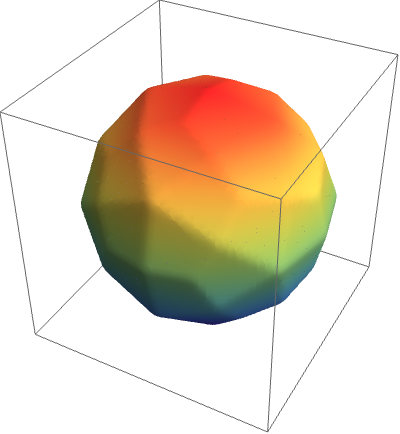}
\end{minipage}
\begin{minipage}[b]{0.16\hsize}
\centering
\includegraphics[width=0.99\textwidth]{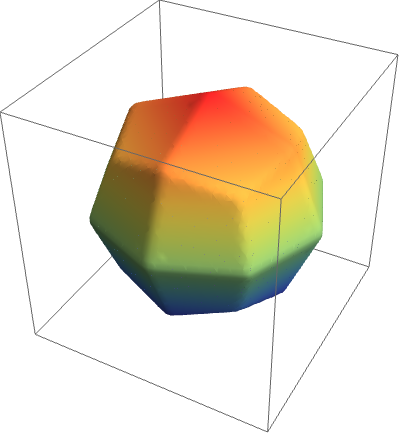}
\end{minipage}
\centering
\begin{minipage}[b]{0.16\hsize}
\centering
\includegraphics[width=0.99\textwidth]{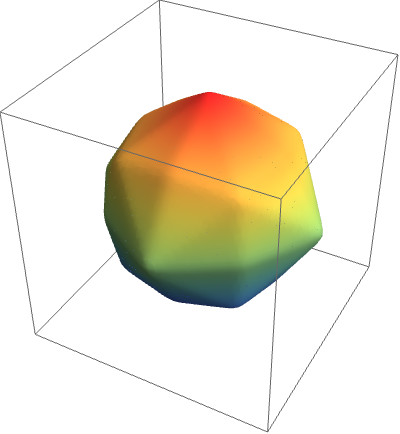}
\end{minipage}
\begin{minipage}[b]{0.16\hsize}
\centering
\includegraphics[width=0.99\textwidth]{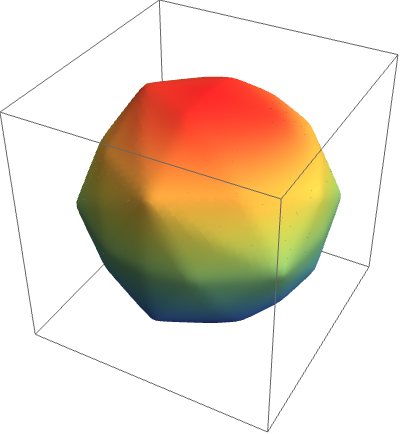}
\end{minipage}
\begin{minipage}[b]{0.16\hsize}
\centering
\includegraphics[width=0.99\textwidth]{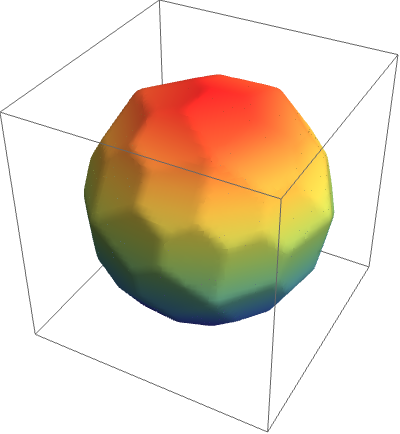}
\end{minipage}
\begin{minipage}[b]{0.16\hsize}
\centering
\includegraphics[width=0.99\textwidth]{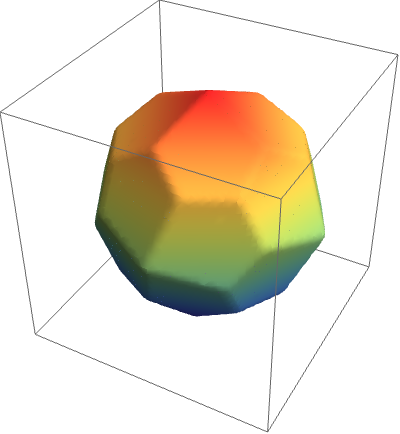}
\end{minipage}
\begin{minipage}[b]{0.16\hsize}
\centering
\includegraphics[width=0.99\textwidth]{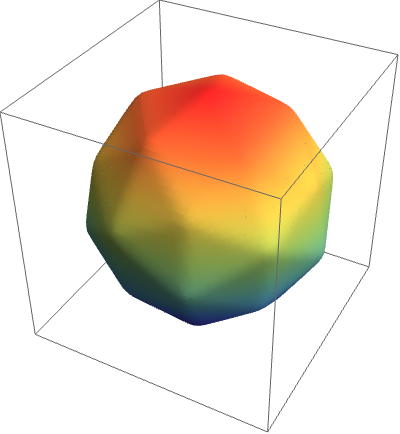}
\end{minipage}
\centering
\begin{minipage}[b]{0.16\hsize}
\centering
\includegraphics[width=0.99\textwidth]{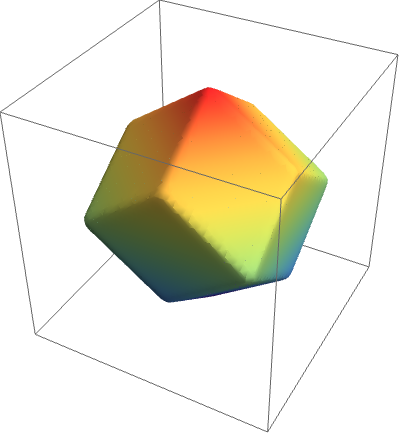}
\end{minipage}
\begin{minipage}[b]{0.16\hsize}
\centering
\includegraphics[width=0.99\textwidth]{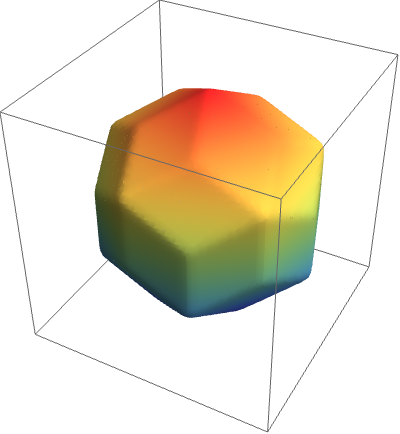}
\end{minipage}
\begin{minipage}[b]{0.16\hsize}
\centering
\includegraphics[width=0.99\textwidth]{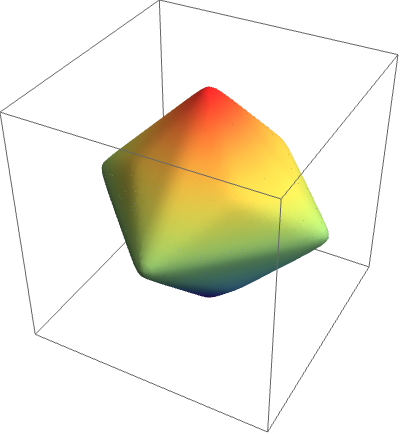}
\end{minipage}
\begin{minipage}[b]{0.16\hsize}
\centering
\includegraphics[width=0.99\textwidth]{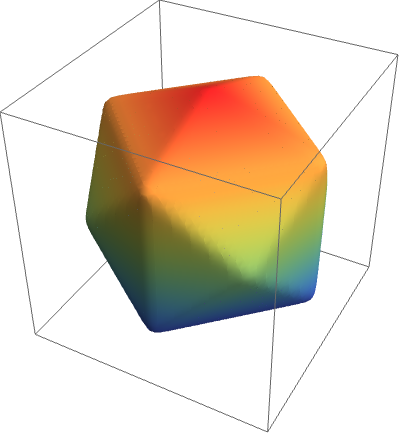}
\end{minipage}
\begin{minipage}[b]{0.16\hsize}
\centering
\includegraphics[width=0.99\textwidth]{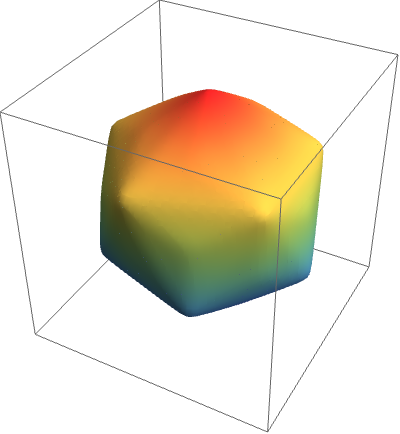}
\end{minipage}
\centering
\begin{minipage}[b]{0.16\hsize}
\centering
\includegraphics[width=0.99\textwidth]{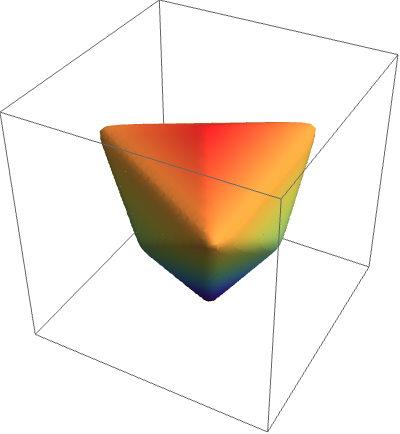}
\end{minipage}
\begin{minipage}[b]{0.16\hsize}
\centering
\includegraphics[width=0.99\textwidth]{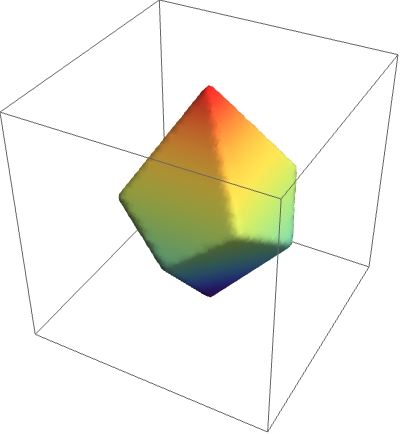}
\end{minipage}
\begin{minipage}[b]{0.16\hsize}
\centering
\includegraphics[width=0.99\textwidth]{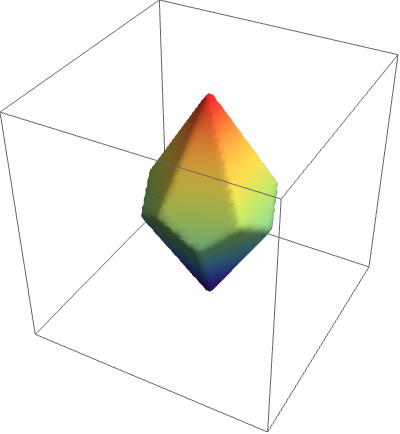}
\end{minipage}
\begin{minipage}[b]{0.16\hsize}
\centering
\includegraphics[width=0.99\textwidth]{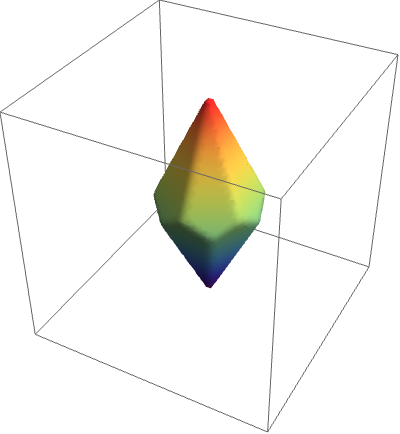}
\end{minipage}
\begin{minipage}[b]{0.16\hsize}
\centering
\includegraphics[width=0.99\textwidth]{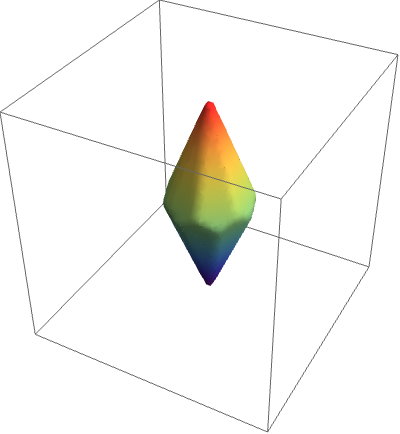}
\end{minipage}
\caption{The massless polyhedral vacuum fermions. The first five are regular polyhedrons called Platonic polyhedra. The next thirteen are the semiregular polyhedrons called the Archimedean polyhedra. Next thirteens are dual of the Archimedean polyhedra, and the last fours are trapezohedrons.}
\label{fig:fermion_polyhedron}
\end{figure}

We have shown that the fermion zero modes can be localized any solitons: monopoles, strings, and domain walls.
These are called the monopole fermions, string fermions, and the domain wall fermions, respectively.
In addition, we also showed that they can be stored in the finite vacuum regions like the cube or octahedron in Fig.~\ref{fig:3d_polyhedron_fermion}.
This is a natural extension of the two dimensional version found in Ref.~\cite{Eto:2023orr}.
By modifying the background soliton, we can localize the fermion zero modes in regions of various shapes, including regular polyhedra and semiregular polyhedra etc as shown in Fig.~\ref{fig:fermion_polyhedron}. We call them the massless polyhedral vacuum fermions.
Hence, our model provides the simplest model for confining massless fermions in a three-dimensional box of arbitrary shape.

\subsection{The localization position of the fermion zero mode}
\label{sec:massshift_fermion}

In the previous section, we observed that the localized points of the fermion zero modes in real space vary in various ways
according to the mass vectors $\vec m_A$ as shown in Figs.~\ref{fig:3d_4vac_fermion}, \ref{fig:3d_5vac_fermion}, and \ref{fig:3d_polyhedron_fermion}. Here we explain where the fermion zero mode functions have nonzero support and how to change the localization position. Let us first look at the fermion mass matrix
\be
M_{\rm f} = \frac{h}{2} \sum_m \varphi_m \sigma^m = \left(\frac{h}{4} \sum_m  s_m  \p_m \log\Omega\right) \sigma^m\,.
\ee
This is in general the function of $(x,y,z)$ for inhomogeneous background $\vec\varphi(x)$.
The zero mode is localized around a massless point where the rank of $M$ reduces:
\be
- \det M_{\rm f} = \frac{h^2}{4} \sum_m  \varphi_m^2 = \frac{h^2}{16} \sum_m  \left(\p_m\log \Omega\right)^2 = 0\,.
\ee
Namely, $\det M_{\rm f} = 0$ occurs at $\vec \varphi = 0$.
Therefore, the localization point is an inverse image of the map $\vec x \to \vec \varphi = 0$.
Indeed, the previous examples in Figs.~\ref{fig:3d_4vac_fermion}, \ref{fig:3d_5vac_fermion}, and \ref{fig:3d_polyhedron_fermion} can be well explained by this way.

Here we can recall the comment around Eq.~(\ref{eq:phi_shift}) where we said that the constant shift of $\vec \varphi$ is unphysical since it can be absorbed by the constant shift of the boson mass $\vec m_A \to \vec m_A - \vec m_0$.
This implies that the origin of $\vec\varphi$ is unphysical.
However, this is only true in the bosonic model without fermions. 
When fermions are included, the origin of the internal space $\vec\varphi$
acquires the physical meaning of being a localized point of massless fermions.
Conversely, the boson mass shift  induces shift of the fermion zero mode 
whereas the bosonic background configuration being unchanged.
This can be explained in slightly different way: Instead of shifting boson mass, we just add the fermion bulk mass in Eq.~(\ref{eq:fermionLagrangian}) as
\be
\mathcal{L}_\mathrm{F}=i\bar{\Psi}_a\gamma^\mu\partial_\mu\Psi_a-h\bar{\Psi}_a 
\frac{(\vec \sigma)_{ab}}{2}\cdot (\vec \varphi - \vec m_0) \Psi_b\,.
\ee
This shifts the zero of $\det M_{\rm f}$ from  $\vec \varphi = 0$ to $\vec \varphi = \vec m_0$.
Of course, this is same as the constant shift of boson mass.

Whether we shift the boson masses or add the fermion bulk mass, we arrive at the same conclusion.
The change in $\Omega$ is given in Eq. (\ref{eq:omega_shift}), and therefore the normalizable fermion zero mode function $\Omega^{-h/4}$
just gets multiplied by the following exponential factor
\begin{gather}
\Omega^{-\frac{h}{4}} 
\rightarrow 
\exp\left(-\frac{h}{2}\sum_m s_m m_{m,0}x^m\right) \Omega^{-\frac{h}{4}} \,.
\end{gather}

\begin{figure}[htbp]
\centering
\begin{minipage}[b]{0.32\hsize}
\centering
\includegraphics[width=0.9\textwidth]{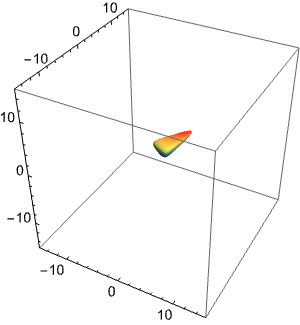}
\subcaption{\(\vec m=(0,m/5,m/5)\)}
\end{minipage}
\begin{minipage}[b]{0.32\hsize}
\centering
\includegraphics[width=0.9\textwidth]{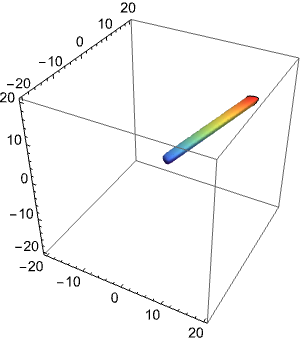}
\subcaption{\(m_{m,0}=\frac{m_{m,1}+m_{m,2}+m_{m,3}}{3}\)}
\end{minipage}
\\
\begin{minipage}[b]{0.32\hsize}
\centering
\includegraphics[width=0.9\textwidth]{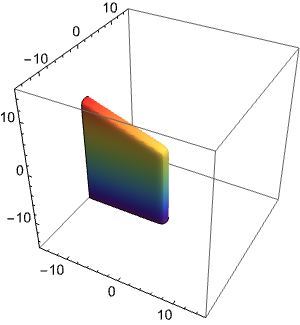}
\subcaption{\(m_{m,0}=\frac{m_{m,3}+m_{m,4}}{2}\)}
\end{minipage}
\centering
\begin{minipage}[b]{0.32\hsize}
\centering
\includegraphics[width=0.9\textwidth]{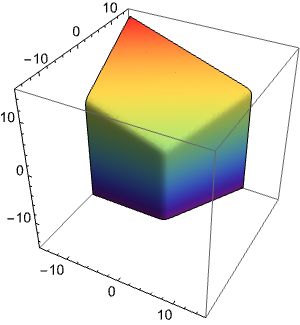}
\subcaption{\(m_{m,0}=m_{m,3}\)}
\end{minipage}
\caption{
The fermion zero mode function \(f\) for the regular tetrahedron case with the mass shift \(m_{m,0}\).
}
\label{fig:3d_fermion_shift}
\end{figure}

Let us illustrate effects of the additional dressing factor 
by focusing on one of the simplest example of the monopole-string-domain wall background in $N_{\rm F} = 4$ with the regular tetrahedral mass arrangement in Eq.~(\ref{eq:massvectors1}). The fermion zero mode for the unshifted case is shown in Fig.~\ref{fig:3d_4vac_fermion}(a)
where the zero mode is localized on the monopole because $\vec\varphi = 0$ is an interior point of the mass tetrahedron.
Now we turn on $\vec m_0$ as the fermion mass shift. The results depend on where $\vec \varphi = \vec m_0$ is located relative to the mass tetrahedron in the $\vec \varphi$ space.
There are five qualitatively different cases.
If \(\vec m_0\) remains inside the mass tetrahedron, the zero mode undergoes a small deformation 
but is still localized around the monopole like Fig.~\ref{fig:3d_fermion_shift}(a).
If \(\vec m_0\) is placed on one of four faces of the tetrahedron, the zero mode 
appears on the corresponding semi-infinite string like 
Fig.~\ref{fig:3d_fermion_shift}(b). 
The zero mode function decreases exponentially rapidly in the transverse direction of the string.
It has a peak at the centre of the string and therefore is not normalizable because it exists everywhere in the semi-infinite string.
However, since it does not diverge, the zero mode can be considered to be localized on a semi-infinite string in the usual sense.
If \(\vec m_0\) is located on one of six edges of the tetrahedron, the zero mode is localized on the corresponding semi-infinite domain wall like 
Fig.~\ref{fig:3d_fermion_shift}(c).
Furthermore, if \(\vec m_0\) is put precisely on one of four vertices of the tetrahedron, the  zero mode emerges at the corresponding vacuum region like Fig.~\ref{fig:3d_fermion_shift}(d).
We consider the zero modes that appear on the semi-infinite domain wall and on the semi-infinite vacuum to be localized in the same sense as those localized on the semi-infinite string.
In contrast, if \(\vec m_0\) is outside of the tetrahedron, the zero mode function is non-normalizable in the sense that it diverges exponentially rapidly. Namely, the fermion zero mode does not exist.

In summary, the volume, faces, edges, and vertices of a generic mass polyhedrons in the \(\varphi_m\) space 
correspond to the monopoles, strings, domain walls, and vacua, respectively, and we can quickly understand 
where the fermion zero mode is localized by just looking at $\vec m_0$.

\subsection{Superconducting current on string-domain wall backgrounds}

Here we restrict ourselves to the background configuration that admits the translational symmetry along the $z$ axis.
This is realized by setting all the third component $m_{3,A}$ ($A=1,\cdots,N_{\rm F}$) be equal to each other.
For simplicity, let us set $m_{3,A} = 0$ in what follows.

In this subsection, we will use the chiral representation 
\be
\tilde \gamma^\mu = 
\left(
\begin{array}{cc}
0 & \sigma^\mu\\
\bar\sigma^\mu & 0
\end{array}
\right)\,,\quad
\tilde\gamma_5 =
\left(
\begin{array}{cc}
1 & 0 \\
0 & -1
\end{array}
\right)\,,\quad
\tilde \Psi_a = 
\left(
\begin{array}{c}
\tilde \chi_a\\
\bar{\tilde\xi}_a
\end{array}
\right)\,.
\ee
The two basis are related by the following unitary transformation $U$
\be
\tilde \gamma^\mu = U \gamma^\mu U^\dag\,,\quad
\tilde \Psi = U \Psi\,,\quad
U = \frac{1}{\sqrt2}\left(
\begin{array}{cc}
1 & -1 \\
-i & - i
\end{array}
\right)\,.
\ee
The Dirac equation in the new basis reads
\begin{gather}
i\tilde\gamma^\mu\partial_\mu\tilde\Psi_a-\frac{h}{2}\sigma_{ab}^m\varphi_m\tilde\Psi_b=0.
\end{gather}
Since we assume $m_{3,A} = 0$, we have $\varphi_3 = 0$ and $\varphi_{1,2}=\varphi_{1,2}(x,y)$.
Then the Dirac equation reduces to
\be
i\tilde\gamma^\mu\p_\mu \tilde\Psi_1 = \frac{h}{2}(\varphi_1 - i \varphi_2)\tilde\Psi_2\,,\\
i\tilde\gamma^\mu\p_\mu \tilde\Psi_2 = \frac{h}{2}(\varphi_1 + i \varphi_2)\tilde\Psi_1\,.
\ee
Let us introduce the complex scalar field by $\phi = \varphi_1 + i\varphi_2$. Then the Dirac equation is expressed as
\be
\left(
\begin{array}{cc}
0 & i\sigma^\mu\p_\mu\\
i\bar\sigma^\mu\p_\mu & 0
\end{array}
\right)
\left(
\begin{array}{c}
\tilde \chi_1\\
\bar{\tilde\xi}_1
\end{array}
\right)
= \frac{h}{2}\phi^*
\left(
\begin{array}{c}
\tilde \chi_2\\
\bar{\tilde\xi}_2
\end{array}
\right)\,,\\
\left(
\begin{array}{cc}
0 & i\sigma^\mu\p_\mu\\
i\bar\sigma^\mu\p_\mu & 0
\end{array}
\right)
\left(
\begin{array}{c}
\tilde \chi_2\\
\bar{\tilde\xi}_2
\end{array}
\right)
= \frac{h}{2}\phi
\left(
\begin{array}{c}
\tilde \chi_1\\
\bar{\tilde\xi}_1
\end{array}
\right)\,.
\ee
Looking carefully the matrix structure, we realize that these can be separated into two independent equations as
\be
\left(
\begin{array}{cc}
0 & i\sigma^\mu\p_\mu\\
i\bar\sigma^\mu\p_\mu & 0
\end{array}
\right)
\left(
\begin{array}{c}
\tilde \chi_1\\
\bar{\tilde\xi}_2
\end{array}
\right)
= \frac{h}{2}
\left(
\begin{array}{cc}
\phi & 0 \\
0 & \phi^*
\end{array}
\right)
\left(
\begin{array}{c}
\tilde \chi_1\\
\bar{\tilde\xi}_2
\end{array}
\right)\,,\label{eq:Deq1_new}\\
\left(
\begin{array}{cc}
0 & i\sigma^\mu\p_\mu\\
i\bar\sigma^\mu\p_\mu & 0
\end{array}
\right)
\left(
\begin{array}{c}
\tilde \chi_2\\
\bar{\tilde\xi}_1
\end{array}
\right)
= \frac{h}{2}
\left(
\begin{array}{cc}
\phi^* & 0 \\
0 & \phi
\end{array}
\right)
\left(
\begin{array}{c}
\tilde \chi_2\\
\bar{\tilde\xi}_1
\end{array}
\right)\,.\label{eq:Deq2_new}
\ee
Next, we separate the variables $x^{0,3}$ and $x^{1,2}$  as 
\begin{gather}
\begin{pmatrix}
\tilde\chi_1(x^\mu) \\
\bar{\tilde\xi}_2(x^\mu)
\end{pmatrix}
=\expon^{-i\omega t + i k z}
\begin{pmatrix}
\tilde\chi_{1}^{(\omega,k)}(x,y) \\
\bar{\tilde\xi}_2^{(\omega,k)}(x,y)
\end{pmatrix}\,,\label{eq:sep_vari_1}\\
\begin{pmatrix}
\tilde\chi_2(x^\mu) \\
\bar{\tilde\xi}_1(x^\mu)
\end{pmatrix}
=\expon^{-i\omega' t + i k' z}
\begin{pmatrix}
\tilde\chi_{2}^{(\omega',k')}(x,y) \\
\bar{\tilde\xi}_1^{(\omega',k')}(x,y)
\end{pmatrix}\,.\label{eq:sep_vari_2}
\end{gather}
Substituting Eq.~(\ref{eq:sep_vari_1}) into (\ref{eq:Deq1_new}), we have
\be
\left(
\begin{array}{cc}
0 & \omega - \sigma^3k +  i\sigma^i\p_i\\
\omega + \sigma^3k - i\sigma^i\p_i & 0
\end{array}
\right)
\left(
\begin{array}{c}
\tilde \chi_1^{(\omega,k)}\\
\bar{\tilde\xi}_2^{(\omega,k)}
\end{array}
\right)
= \frac{h}{2}
\left(
\begin{array}{cc}
\phi & 0 \\
0 & \phi^*
\end{array}
\right)
\left(
\begin{array}{c}
\tilde \chi_1^{(\omega,k)}\\
\bar{\tilde\xi}_2^{(\omega,k)}
\end{array}
\right)\,.
\ee
This can be rewritten as follows
\be
\left(
\begin{array}{cc}
- \sigma^3k + i \sigma^i\p_i & \frac{h}{2}\phi^* \\
\frac{h}{2}\phi & \sigma^3k - i \sigma^i\p_i
\end{array}
\right)
\left(
\begin{array}{c}
\tilde \chi_1^{(\omega,k)}\\
\bar{\tilde\xi}_2^{(\omega,k)}
\end{array}
\right)
=
\omega \left(
\begin{array}{c}
\tilde \chi_1^{(\omega,k)}\\
\bar{\tilde\xi}_2^{(\omega,k)}
\end{array}
\right)\,.
\ee
Defining the following Hermitian operators
\be
H_k = k Z + M\,,\quad
Z = \left(
\begin{array}{cc}
- \sigma^3 & 0 \\
0 & \sigma^3
\end{array}
\right)\,,\quad
M = 
\left(
\begin{array}{cc}
i \sigma^i\p_i & \frac{h}{2}\phi^* \\
\frac{h}{2}\phi & - i \sigma^i\p_i
\end{array}
\right)\,.
\ee
Then Eq.~(\ref{eq:Deq1_new}) ends up with
\be
H_k \left(
\begin{array}{c}
\tilde \chi_1^{(\omega,k)}\\
\bar{\tilde\xi}_2^{(\omega,k)}
\end{array}
\right)
= \omega \left(
\begin{array}{c}
\tilde \chi_1^{(\omega,k)}\\
\bar{\tilde\xi}_2^{(\omega,k)}
\end{array}
\right)\,.\label{eq:H}
\ee
Obviously, Eq.~(\ref{eq:Deq2_new}) can be cast into a similar form as
\be
H_k' \left(
\begin{array}{c}
\tilde \chi_2^{(\omega',k')}\\
\bar{\tilde\xi}_1^{(\omega',k')}
\end{array}
\right)
= \omega' \left(
\begin{array}{c}
\tilde \chi_2^{(\omega',k')}\\
\bar{\tilde\xi}_1^{(\omega',k')}
\end{array}
\right)\,,\label{eq:H'}
\ee
with
\be
H'_{k'} = k' Z + M'\,,\quad
Z = \left(
\begin{array}{cc}
- \sigma^3 & 0 \\
0 & \sigma^3
\end{array}
\right)\,,\quad
M' = 
\left(
\begin{array}{cc}
i \sigma^i\p_i & \frac{h}{2}\phi \\
\frac{h}{2}\phi^* & - i \sigma^i\p_i
\end{array}
\right)\,,
\ee
Since $\{Z,M\} = \{Z,M'\} = 0$, we have
\be
H_k^2 = k^2 + M^2\,,\quad H_{k'}'{}^2 = k'{}^2 + M'{}^2\,.
\ee
Thus, $H_k^2$ ($H_{k'}'{}^2$) commute with $M$ ($M'$), and therefore they are simultaneously diagonalized.
Now, we only need to find out eigenvalues $m_n$ and $m_n'$ of $M$ and $M'$, respectively. Once we get them, $\omega$ and $\omega'$
are given by
\be
\omega_n^2 = k^2 + m_n^2\,,\quad
\omega_n'{}^2 = k'{}^2 + m_n'{}^2\,.
\ee

From now on we will focus on zero modes $m_0 = m_0' = 0$. 
The corresponding eigenvectors for $M$ and $M'$ are decomposed as follows and they satisfy the
eigenvalue equations
\be
M 
\left(
\begin{array}{c}
\tilde g\\
\tilde f\\
i\tilde f^*\\
i\tilde g^*
\end{array}
\right)
= 0\,,\quad
M' 
\left(
\begin{array}{c}
\tilde g'\\
\tilde f'\\
i\tilde f'{}^*\\
i\tilde g'{}^*
\end{array}
\right)
= 0\,.
\ee
The explicit expressions are given by
\be
(\p_1 - i \p_2)\tilde f + \frac{h}{2} \phi^*\tilde f^* = 0\,, \label{eq:zeromode_2d_1}\\
(\p_1 + i \p_2)\tilde g + \frac{h}{2} \phi^* \tilde g^* = 0\,, \label{eq:zeromode_2d_2}\\
(\p_1 - i \p_2)\tilde f' + \frac{h}{2} \phi \tilde f'{}^* = 0\,, \label{eq:zeromode_2d_3}\\
(\p_1 + i \p_2)\tilde g' + \frac{h}{2} \phi  \tilde g'{}^* = 0\,.\label{eq:zeromode_2d_4}
\ee

We now need to fix the background solution $\phi$. Since we chose $m_{3,A} = 0$ ($A=1,2,\cdots,N_{\rm F}$, no fields depend on $x^3$. 
The BPS equations (\ref{eq:BPS1}) and (\ref{eq:BPS2}) do not need to be changed, just replacing $\p_3 \to 0$ is enough. 
The same holds for the solution $\varphi_m$ in Eq.~(\ref{eq:varphi}), namely $\phi = \frac{1}{2}(s_1\p_1 + is_2\p_2)\log\Omega$.
To be concrete, let us take $(s_1,s_2) = (1,1)$ in the following.
Then we find two solutions for each of Eq.~(\ref{eq:zeromode_2d_1}) and (\ref{eq:zeromode_2d_4}) as
\be
\tilde f = \{\Omega^{-\frac{h}{4}}\,, i \Omega^{\frac{h}{4}}\}\,,\quad
\tilde g' = \{\Omega^{-\frac{h}{4}}\,, i \Omega^{\frac{h}{4}}\}\,,
\ee
where as $\tilde g = \tilde f' = 0$ for Eqs.~(\ref{eq:zeromode_2d_2}) and (\ref{eq:zeromode_2d_3}).
Since $\Omega^{\frac{h}{4}}$ is not normalizable, we conclude that there are two independent normalizable zero modes:
\be
\left(
\begin{array}{c}
\tilde \chi_1^{(\omega,k)}\\
\bar{\tilde\xi}_2^{(\omega,k)}
\end{array}
\right)
= \Omega^{-\frac{h}{4}} \left(
\begin{array}{c}
0\\
1\\
i\\
0
\end{array}
\right)\,,\quad
\left(
\begin{array}{c}
\tilde \chi_2^{(\omega',k')}\\
\bar{\tilde\xi}_1^{(\omega',k')}
\end{array}
\right)
= \Omega^{-\frac{h}{4}} \left(
\begin{array}{c}
1\\
0\\
0\\
i
\end{array}
\right)\,.
\ee
Finally, let us determine the dispersion relation.
The left (right) one is the eigenstate of $Z$ with eigenvalue $1$ ($-1$).
This together with Eqs.~(\ref{eq:H}) and (\ref{eq:H'}) gives us $\omega = k$ for the first zero mode solution while
$\omega' = - k'$ for the second zero mode solution.
Thus, the fermion zero modes including $t$ and $z$ are given by
\begin{gather}
\begin{pmatrix}
\tilde\chi_1^{(0)} \\
\bar{\tilde\xi}_2^{(0)}
\end{pmatrix}
=\expon^{-ik (t - z)}
\Omega^{-\frac{h}{4}} \left(
\begin{array}{c}
0\\
1\\
i\\
0
\end{array}
\right)\,,\quad
\begin{pmatrix}
\tilde\chi_2^{(0)} \\
\bar{\tilde\xi}_1^{(0)}
\end{pmatrix}
=\expon^{i k' (t + z)}
\Omega^{-\frac{h}{4}} \left(
\begin{array}{c}
1\\
0\\
0\\
i
\end{array}
\right)
\,.\label{eq:zero_modes_2d}
\end{gather}
The left one corresponds to a massless chiral fermion moving at the speed of light in the positive $z$-direction (right-moving by convention). 
The right one corresponds to a massless chiral fermion moving at the speed of light in the negative $z$-direction (left-moving by convention). 
The two dimensional ($tz$) chirality can be also understood from the chirality matrix $\gamma_0\gamma_3$ in the two dimensions. 
The first solution of Eq.~(\ref{eq:zero_modes_2d}) is the eigenstate of $\gamma_0\gamma_3$ with the eigenvalue $+1$ whereas
the second one has the eigenvalue $-1$. Thus, the low-energy effective theory of the world volume of the string-domain wall composite is a $1+1$-dimensional theory with both the left-moving and the right-moving chiral fermions. This is a vector-like theory and therefore not anomalous.
Hence, no anomaly inflow exists from the bulk to the  string-domain wall world volume. 
Recently, we studied the string-domain wall composites with chiral superconducting current in Ref.~\cite{Eto:2023orr}. 
In Ref.~\cite{Eto:2023orr} the number of fermions is a half of that in this paper, namely, $\tilde \chi_2$ and $\bar{\tilde\xi}_1$ are absent.
Hence, the only right-moving current exists and flows on the string-domain walls. In contrast, the vector-like current flows 
on the string-domain wall along the $z$ axis in this paper.

For the paper to be self contained, let us also show the zero modes in the previous basis without tilde. They are given by
\be
\Psi_1
&=& 
\Omega^{-\frac{h}{4}}
\left(
\begin{array}{c}
0\\
c e^{ik(-t+z)} - c' e^{ik'(t+z)}\\
0\\
- c e^{ik(-t+z)} - c' e^{ik'(t+z)}
\end{array}
\right)\,, \label{eq:zeromode_2dim_1}\\
\Psi_2
&=& 
\Omega^{-\frac{h}{4}}
\left(
\begin{array}{c}
- c e^{ik(-t+z)} + c' e^{ik'(t+z)}\\
0\\
- c e^{ik(-t+z)} - c' e^{ik'(t+z)}\\
0
\end{array}
\right)\,,
\label{eq:zeromode_2dim_2}
\ee
where $c$ and $c'$ are real constants.
Let us compare this with the previously obtained zero modes for the monopole-string-domain wall in Sec.~\ref{sec:fermion_zero_mode}.
First of all, we realize difference of the number of zero modes. We found one fermion zero mode for the monopole-string-domain wall for the specific choice of the sign combination $\vec s$ in Sec.~\ref{sec:fermion_zero_mode}, whereas we have two independent zero modes
corresponding to $(c,c') = (1,0)$ and $(0,1)$ in Eqs.~(\ref{eq:zeromode_2dim_1}) and (\ref{eq:zeromode_2dim_2}).
To solve this puzzle, we recall that 
the fermion zero modes with non-zero $k$ ($k'$) is only possible for the string-domain wall composite background 
that has the translational symmetry along the $z$ axis. 
Indeed, all the calculations in this subsection are valid only for the background fields $\varphi_{1,2} = \varphi_{1,2}(x,y)$ and $\varphi_3 = 0$
that do not depend on $z$. For generic three dimensional composites of the monopole-string-domain wall,
there is no room for the zero mode wave function to have the dependence ${\rm e}^{ikz}$. 
Indeed,  the fermion zero mode functions for the monopole-string-domain walls
are always localized on a (semi) finite region, like around a monopole, a finite (semi-infinite) string or a finite (semi-infinite) domain wall.
If a fermion current could flow, it should be conserved. But there are no directions along which the fermion zero mode comes into and goes out for the monopole-string-domain wall fermions. Hence, the current cannot flow on the monopole-string-domain wall background. 
Therefore, we should compare the zero modes of the monopole-string-domain wall background and those of the string-domain wall 
backgrounds with $k = k' = 0$. Indeed, we find that the latter can be expressed as linear combinations of the former as
\be
\Psi_1\big|_{k=k'=0} 
&=& 
(c-c')\Psi_1^{(+,+,+)} - (c + c')\Psi_1^{(+,+,-)}
\,,\\
\Psi_2\big|_{k=k'=0} 
&=& 
(c-c')\Psi_2^{(+,+,+)} - (c + c')\Psi_2^{(+,+,-)}\,,
\ee
where $\Psi_{1,2}$ on the lefthand side is the zero mode of the string-domain wall given in Eqs.~(\ref{eq:zeromode_2dim_1})
and (\ref{eq:zeromode_2dim_2}) with $k=k'=0$ while $\Psi^{(+++)}$ and $\Psi^{(++-)}$ on the righthand side are the zero modes
of the monopole-string-domain wall with $\vec s = (1,1,1)$ and $(1,1,-1)$ given in Eqs.~(\ref{eq:+++}) and (\ref{eq:++-}).
This clearly explains why there are two massless fermions for the string-domain wall background: When the background configuration
does not depend on $z$, the third component $s^3$ does not appear. Then both $s^3 = 1$ and $-1$ should be taken into account,
providing two massless zero modes.

\section{Summaries and discussions}
\label{sec:summary}

In this paper, we studied massless fermions localized on the three dimensional monopole-string-domain wall composites.
We proved the existence of the single fermion zero mode localized on the composite solitons by analytically solving the Dirac equation.
We found all analytical solutions are generally proportional to $\Omega^{-h/4}$, where \(\Omega\) is the gauge invariant scalar function introduced by the moduli matrix method.
We then concretely demonstrated that massless fermions can appear on monopoles, on finite or semi-infinite strings, on finite or semi-infinite domain walls, or in semi-infinite or finite polyhedral vacua. We showed in which part of the network the massless fermion is localized can be understood from the singular point $(\vec\varphi = \vec m_0$) of the mass matrix $M_{\rm f}$ of the fermion.
When it occurs on one of the vertices, one of the edges, one of the faces or inside the body of the mass polyhedron in the \(\vec\varphi\) space, the massless fermion appears on the corresponding vacuum, domain wall, string or monopole, respectively.
In addition, we discovered ``massless polyhedral vacuum fermions", which are massless fermions confined within the vacuum regions in the form of various convex polyhedra.
If the singular point lies outside the mass polyhedron, then there are no massless fermions that can be normalized.
Furthermore, we studied the superconducting current on the monopole-string-domain walls composites.
In general, if a composite soliton consists of monopoles, strings, and domain walls and spans three spatial dimensions without translational symmetry, no current will flow.
On the other hand, if the soliton network consists only of strings and domain walls and has translational symmetry about the $z$-axis, then both left-handed and right-handed massless fermions are localized. The former (the latter) propagates in the positive (negative) \(z\) direction, resulting in a vector-like superconducting current.
This is in stark contrast to the chiral current found in our previous work~\cite{Eto:2023orr}.

Let us make several comments on possible future directions.
In this paper, we only dealt with the $SU(2)$ isospinor fermions $\Psi_{a=1,2}$ coupled to monopole-string-domain wall composites by the Yukawa term in Eq.~(\ref{eq:fermionLagrangian}), and found there exists the single zero mode. 
It turns out that in Ref.~\cite{Jackiw:1975fn} the $SU(2)$ isovecotor fermions $\Psi_{\alpha = 1,2,3}$ under a spherical monopole background have two normalizable zero modes, twice as many as isospinor fermions. Unfortunately, in this paper we do not answer the elementary question of how many zero modes exist for the isovector fermions coupled to the monopole-string-domain wall composites.
One obstacle is that the Dirac equation, unlike the isospinor case, is difficult to solve analytically.
It is not yet clear whether an analytical solution can be obtained, but it may be possible to calculate the number of zero modes using the index theorem \cite{Weinberg:1981eu,Ganoulis:1987np}.
The index theorems for individual solitons such as single domain walls and strings are well established, but its application to composite solitons has not yet been explored well. This issue remains to be addressed in the future.
Another direction to go is studying fermionic zero modes in more general composite solitons.
In this paper we only consider the BPS monopole-string-domain wall composites in the Abelian-Higgs model in three spatial dimensions. 
The generalization of these composites to a Yang-Mills-Higgs model in higher dimensions has been found in Ref.~\cite{Eto:2020vjm,Eto:2020cys}, and we would like to investigate whether the fermionic zero modes discovered in this paper also exist in these non-Abelian and higher dimensional composite solitons.
Furthermore, different types of BPS composites are known in ${\cal N}=2$ supersymmetric Yang-Mills-Higgs model~\cite{Isozumi:2004vg}: monopole-string-domain wall, where the string and domain wall are perpendicular to each other. 
Because the moduli matrix formalism also works for these composites, we naively expect a similar analytic solution like $\Omega^{-h/4}$ found in this paper would exist. We will report on this elsewhere.
Finally, in this paper we only focused on the zero modes.
It will be important to understand the entire spectrum, including massive modes, because
the massive modes would affect on the background solitons.
Recently, it is reported that massive fermion modes localized on kinks
can significantly modifies the host soliton and breaks the symmetry between 
the localized modes with positive and negative eigenvalues in Ref.~\cite{Klimashonok:2019iya}.

\appendix
\section{Fermion zero mode for a spherically symmetric global monopole}
\label{sec:appA}

Let us give a quick review on the well-known fermion zero modes on a spherically symmetric monopole.
Suppose $\varphi_m$ takes a spherically symmetric configuration 
\be
\varphi_m = F(r) \frac{x_m}{r}\,,
\ee
with $r = |\vec x|$. $F(r)$ is the profile function which takes $0$ at $r=0$ and asymptotically goes to a constant as $F \to \mu$ as $r \to \infty$.
This configuration is called the global monopole. This is, of course, not a solution of our model (\ref{eq:Lagrangian_boson}), but we here consider
a usual $\varphi^4$ theory like
\be
{\cal L} = \frac{1}{2}\sum_m \p_\mu \varphi_m \p^\mu \varphi_m - \frac{\lambda}{4}\left(\sum_m\varphi_m^2 - \mu^2\right)^2\,.
\ee
The monopole charge density ${\cal M}$ given in Eq.~(\ref{eq:M_density}) reads
\be
{\cal M} = \frac{F'F^2}{r^2}\,,
\ee
and the monopole charge $M$ given in Eq.~(\ref{eq:M}) is explicitly calculated as 
\be
M = \int {\cal M}\, d^3x = 4\pi \int^\infty_0 F'F^2\, dr = \frac{4\pi \mu^3}{3}\,.
\ee
This corresponds to the volume of vacuum sphere of the radius $\mu$.

Let us find out a zero mode of the Dirac equation (\ref{eq:Diraceq}) for the global monopole as a background configuration.
The appropriate ansatz \cite{Jackiw:1975fn,Weinberg:2012pjx} is known as
\begin{gather}
\chi_{a\alpha}^{(0)}=f(r)\epsilon_{a\alpha}
,\quad \bar{\xi}_a^{\dot{\alpha}(0)}=g(r)\epsilon^{a\dot{\alpha}}.
\label{eq:ansatz_fermion_single}
\end{gather}
with the profile functions $f(r)$ and $g(r)$ depending on $r$ only.
The Dirac equation reduces to
\be
f' + \frac{h}{2}Ff=0\,,\quad
g' - \frac{h}{2}Fg=0\,.
\ee
Therefore, there exist one normalizable zero mode
\be
f(r) = \exp\left(-\frac{h}{2}\int^r F(r')\, dr'\right)\,,\quad g(r) = 0\,.
\ee

\begin{acknowledgments}

This work is supported in part by JSPS KAKENHI Grants No. JP22H01221 (ME).

\end{acknowledgments}

\bibliographystyle{jhep}
\normalem
\bibliography{references}

\providecommand{\href}[2]{#2}\begingroup\raggedright\begin{thebibliography}{10}

\bibitem{Abrikosov:1956sx}
A.~A. Abrikosov, \emph{{On the Magnetic properties of superconductors of the second group}}, {\emph{Sov. Phys. JETP} {\bfseries 5} (1957) 1174}.

\bibitem{Nielsen:1973cs}
H.~B. Nielsen and P.~Olesen, \emph{{Vortex Line Models for Dual Strings}}, \href{https://doi.org/10.1016/0550-3213(73)90350-7}{\emph{Nucl. Phys. B} {\bfseries 61} (1973) 45}.

\bibitem{tHooft:1974kcl}
G.~'t~Hooft, \emph{{Magnetic Monopoles in Unified Gauge Theories}}, \href{https://doi.org/10.1016/0550-3213(74)90486-6}{\emph{Nucl. Phys. B} {\bfseries 79} (1974) 276}.

\bibitem{Polyakov:1974ek}
A.~M. Polyakov, \emph{{Particle Spectrum in Quantum Field Theory}}, {\emph{JETP Lett.} {\bfseries 20} (1974) 194}.

\bibitem{Belavin:1975fg}
A.~A. Belavin, A.~M. Polyakov, A.~S. Schwartz and Y.~S. Tyupkin, \emph{{Pseudoparticle Solutions of the Yang-Mills Equations}}, \href{https://doi.org/10.1016/0370-2693(75)90163-X}{\emph{Phys. Lett. B} {\bfseries 59} (1975) 85}.

\bibitem{Preskill:1992ck}
J.~Preskill and A.~Vilenkin, \emph{{Decay of metastable topological defects}}, \href{https://doi.org/10.1103/PhysRevD.47.2324}{\emph{Phys. Rev. D} {\bfseries 47} (1993) 2324} [\href{https://arxiv.org/abs/hep-ph/9209210}{{\ttfamily hep-ph/9209210}}].

\bibitem{Eto:2023gfn}
M.~Eto, Y.~Hamada and M.~Nitta, \emph{{Composite topological solitons consisting of domain walls, strings, and monopoles in O(N) models}}, \href{https://doi.org/10.1007/JHEP08(2023)150}{\emph{JHEP} {\bfseries 08} (2023) 150} [\href{https://arxiv.org/abs/2304.14143}{{\ttfamily 2304.14143}}].

\bibitem{Jackiw:1975fn}
R.~Jackiw and C.~Rebbi, \emph{{Solitons with Fermion Number 1/2}}, \href{https://doi.org/10.1103/PhysRevD.13.3398}{\emph{Phys. Rev. D} {\bfseries 13} (1976) 3398}.

\bibitem{Jackiw:1981ee}
R.~Jackiw and P.~Rossi, \emph{{Zero Modes of the Vortex - Fermion System}}, \href{https://doi.org/10.1016/0550-3213(81)90044-4}{\emph{Nucl. Phys. B} {\bfseries 190} (1981) 681}.

\bibitem{WITTEN1985557}
E.~Witten, \emph{Superconducting strings}, \href{https://doi.org/https://doi.org/10.1016/0550-3213(85)90022-7}{\emph{Nuclear Physics B} {\bfseries 249} (1985) 557}.

\bibitem{Peccei:1977hh}
R.~D. Peccei and H.~R. Quinn, \emph{{CP Conservation in the Presence of Instantons}}, \href{https://doi.org/10.1103/PhysRevLett.38.1440}{\emph{Phys. Rev. Lett.} {\bfseries 38} (1977) 1440}.

\bibitem{Weinberg:1977ma}
S.~Weinberg, \emph{{A New Light Boson?}}, \href{https://doi.org/10.1103/PhysRevLett.40.223}{\emph{Phys. Rev. Lett.} {\bfseries 40} (1978) 223}.

\bibitem{Wilczek:1977pj}
F.~Wilczek, \emph{{Problem of Strong $P$ and $T$ Invariance in the Presence of Instantons}}, \href{https://doi.org/10.1103/PhysRevLett.40.279}{\emph{Phys. Rev. Lett.} {\bfseries 40} (1978) 279}.

\bibitem{Preskill:1982cy}
J.~Preskill, M.~B. Wise and F.~Wilczek, \emph{{Cosmology of the Invisible Axion}}, \href{https://doi.org/10.1016/0370-2693(83)90637-8}{\emph{Phys. Lett. B} {\bfseries 120} (1983) 127}.

\bibitem{Abbott:1982af}
L.~F. Abbott and P.~Sikivie, \emph{{A Cosmological Bound on the Invisible Axion}}, \href{https://doi.org/10.1016/0370-2693(83)90638-X}{\emph{Phys. Lett. B} {\bfseries 120} (1983) 133}.

\bibitem{Dine:1982ah}
M.~Dine and W.~Fischler, \emph{{The Not So Harmless Axion}}, \href{https://doi.org/10.1016/0370-2693(83)90639-1}{\emph{Phys. Lett. B} {\bfseries 120} (1983) 137}.

\bibitem{Kibble:1976sj}
T.~W.~B. Kibble, \emph{{Topology of Cosmic Domains and Strings}}, \href{https://doi.org/10.1088/0305-4470/9/8/029}{\emph{J. Phys. A} {\bfseries 9} (1976) 1387}.

\bibitem{Zurek:1985qw}
W.~H. Zurek, \emph{{Cosmological Experiments in Superfluid Helium?}}, \href{https://doi.org/10.1038/317505a0}{\emph{Nature} {\bfseries 317} (1985) 505}.

\bibitem{Davis:1986xc}
R.~L. Davis, \emph{{Cosmic Axions from Cosmic Strings}}, \href{https://doi.org/10.1016/0370-2693(86)90300-X}{\emph{Phys. Lett. B} {\bfseries 180} (1986) 225}.

\bibitem{Lazarides:1984zq}
G.~Lazarides and Q.~Shafi, \emph{{Superconducting Strings in Axion Models}}, \href{https://doi.org/10.1016/0370-2693(85)91398-X}{\emph{Phys. Lett. B} {\bfseries 151} (1985) 123}.

\bibitem{Chudnovsky:1986hc}
E.~M. Chudnovsky, G.~B. Field, D.~N. Spergel and A.~Vilenkin, \emph{{SUPERCONDUCTING COSMIC STRINGS}}, \href{https://doi.org/10.1103/PhysRevD.34.944}{\emph{Phys. Rev. D} {\bfseries 34} (1986) 944}.

\bibitem{Ostriker:1986xc}
J.~P. Ostriker, A.~C. Thompson and E.~Witten, \emph{{Cosmological Effects of Superconducting Strings}}, \href{https://doi.org/10.1016/0370-2693(86)90301-1}{\emph{Phys. Lett. B} {\bfseries 180} (1986) 231}.

\bibitem{Copeland:1987th}
E.~J. Copeland, N.~Turok and M.~Hindmarsh, \emph{{Dynamics of Superconducting Cosmic Strings}}, \href{https://doi.org/10.1103/PhysRevLett.58.1910}{\emph{Phys. Rev. Lett.} {\bfseries 58} (1987) 1910}.

\bibitem{Nielsen:1987fy}
N.~K. Nielsen and P.~Olesen, \emph{{Dynamical Properties of Superconducting Cosmic Strings}}, \href{https://doi.org/10.1016/0550-3213(87)90498-6}{\emph{Nucl. Phys. B} {\bfseries 291} (1987) 829}.

\bibitem{Davis:1988jp}
R.~L. Davis and E.~P.~S. Shellard, \emph{{The Physics of Vortex Superconductivity}}, \href{https://doi.org/10.1016/0370-2693(88)90673-9}{\emph{Phys. Lett. B} {\bfseries 207} (1988) 404}.

\bibitem{CALLAN1985427}
C.~Callan and J.~Harvey, \emph{Anomalies and fermion zero modes on strings and domain walls}, \href{https://doi.org/https://doi.org/10.1016/0550-3213(85)90489-4}{\emph{Nuclear Physics B} {\bfseries 250} (1985) 427}.

\bibitem{Carter:1993wu}
B.~Carter and X.~Martin, \emph{{Dynamic instability criterion for circular (Vorton) string loops}}, \href{https://doi.org/10.1006/aphy.1993.1078}{\emph{Annals Phys.} {\bfseries 227} (1993) 151} [\href{https://arxiv.org/abs/hep-th/0306111}{{\ttfamily hep-th/0306111}}].

\bibitem{Brandenberger:1996zp}
R.~H. Brandenberger, B.~Carter, A.-C. Davis and M.~Trodden, \emph{{Cosmic vortons and particle physics constraints}}, \href{https://doi.org/10.1103/PhysRevD.54.6059}{\emph{Phys. Rev. D} {\bfseries 54} (1996) 6059} [\href{https://arxiv.org/abs/hep-ph/9605382}{{\ttfamily hep-ph/9605382}}].

\bibitem{Martins:1998gb}
C.~J. A.~P. Martins and E.~P.~S. Shellard, \emph{{Vorton formation}}, \href{https://doi.org/10.1103/PhysRevD.57.7155}{\emph{Phys. Rev. D} {\bfseries 57} (1998) 7155} [\href{https://arxiv.org/abs/hep-ph/9804378}{{\ttfamily hep-ph/9804378}}].

\bibitem{Martins:1998th}
C.~J. A.~P. Martins and E.~P.~S. Shellard, \emph{{Limits on cosmic chiral vortons}}, \href{https://doi.org/10.1016/S0370-2693(98)01466-X}{\emph{Phys. Lett. B} {\bfseries 445} (1998) 43} [\href{https://arxiv.org/abs/hep-ph/9806480}{{\ttfamily hep-ph/9806480}}].

\bibitem{Carter:1999an}
B.~Carter and A.-C. Davis, \emph{{Chiral vortons and cosmological constraints on particle physics}}, \href{https://doi.org/10.1103/PhysRevD.61.123501}{\emph{Phys. Rev. D} {\bfseries 61} (2000) 123501} [\href{https://arxiv.org/abs/hep-ph/9910560}{{\ttfamily hep-ph/9910560}}].

\bibitem{Fukuda:2020kym}
H.~Fukuda, A.~V. Manohar, H.~Murayama and O.~Telem, \emph{{Axion strings are superconducting}}, \href{https://doi.org/10.1007/JHEP06(2021)052}{\emph{JHEP} {\bfseries 06} (2021) 052} [\href{https://arxiv.org/abs/2010.02763}{{\ttfamily 2010.02763}}].

\bibitem{Abe:2020ure}
Y.~Abe, Y.~Hamada and K.~Yoshioka, \emph{{Electroweak axion string and superconductivity}}, \href{https://doi.org/10.1007/JHEP06(2021)172}{\emph{JHEP} {\bfseries 06} (2021) 172} [\href{https://arxiv.org/abs/2010.02834}{{\ttfamily 2010.02834}}].

\bibitem{Agrawal:2020euj}
P.~Agrawal, A.~Hook, J.~Huang and G.~Marques-Tavares, \emph{{Axion string signatures: a cosmological plasma collider}}, \href{https://doi.org/10.1007/JHEP01(2022)103}{\emph{JHEP} {\bfseries 01} (2022) 103} [\href{https://arxiv.org/abs/2010.15848}{{\ttfamily 2010.15848}}].

\bibitem{Ibe:2021ctf}
M.~Ibe, S.~Kobayashi, Y.~Nakayama and S.~Shirai, \emph{{On Stability of Fermionic Superconducting Current in Cosmic String}}, \href{https://doi.org/10.1007/JHEP05(2021)217}{\emph{JHEP} {\bfseries 05} (2021) 217} [\href{https://arxiv.org/abs/2102.05412}{{\ttfamily 2102.05412}}].

\bibitem{Kim:1979if}
J.~E. Kim, \emph{{Weak Interaction Singlet and Strong CP Invariance}}, \href{https://doi.org/10.1103/PhysRevLett.43.103}{\emph{Phys. Rev. Lett.} {\bfseries 43} (1979) 103}.

\bibitem{Shifman:1979if}
M.~A. Shifman, A.~I. Vainshtein and V.~I. Zakharov, \emph{{Can Confinement Ensure Natural CP Invariance of Strong Interactions?}}, \href{https://doi.org/10.1016/0550-3213(80)90209-6}{\emph{Nucl. Phys. B} {\bfseries 166} (1980) 493}.

\bibitem{Hiramatsu:2010yn}
T.~Hiramatsu, M.~Kawasaki and K.~Saikawa, \emph{{Evolution of String-Wall Networks and Axionic Domain Wall Problem}}, \href{https://doi.org/10.1088/1475-7516/2011/08/030}{\emph{JCAP} {\bfseries 08} (2011) 030} [\href{https://arxiv.org/abs/1012.4558}{{\ttfamily 1012.4558}}].

\bibitem{Hiramatsu:2012gg}
T.~Hiramatsu, M.~Kawasaki, K.~Saikawa and T.~Sekiguchi, \emph{{Production of dark matter axions from collapse of string-wall systems}}, \href{https://doi.org/10.1103/PhysRevD.85.105020}{\emph{Phys. Rev. D} {\bfseries 85} (2012) 105020} [\href{https://arxiv.org/abs/1202.5851}{{\ttfamily 1202.5851}}].

\bibitem{Gorghetto:2018myk}
M.~Gorghetto, E.~Hardy and G.~Villadoro, \emph{{Axions from Strings: the Attractive Solution}}, \href{https://doi.org/10.1007/JHEP07(2018)151}{\emph{JHEP} {\bfseries 07} (2018) 151} [\href{https://arxiv.org/abs/1806.04677}{{\ttfamily 1806.04677}}].

\bibitem{Gorghetto:2020qws}
M.~Gorghetto, E.~Hardy and G.~Villadoro, \emph{{More axions from strings}}, \href{https://doi.org/10.21468/SciPostPhys.10.2.050}{\emph{SciPost Phys.} {\bfseries 10} (2021) 050} [\href{https://arxiv.org/abs/2007.04990}{{\ttfamily 2007.04990}}].

\bibitem{Eto:2023orr}
M.~Eto and Y.~Suzuki, \emph{{Massless fermions and superconductivity of string-wall composites}}, \href{https://doi.org/10.1007/JHEP02(2024)030}{\emph{JHEP} {\bfseries 02} (2024) 030} [\href{https://arxiv.org/abs/2311.15805}{{\ttfamily 2311.15805}}].

\bibitem{Ito:2000zf}
K.~Ito, M.~Naganuma, H.~Oda and N.~Sakai, \emph{{Nonnormalizable zero modes on BPS junctions}}, \href{https://doi.org/10.1016/S0550-3213(00)00436-3}{\emph{Nucl. Phys. B} {\bfseries 586} (2000) 231} [\href{https://arxiv.org/abs/hep-th/0004188}{{\ttfamily hep-th/0004188}}].

\bibitem{Eto:2020vjm}
M.~Eto, M.~Kawaguchi, M.~Nitta and R.~Sasaki, \emph{{Exact solutions of domain wall junctions in arbitrary dimensions}}, \href{https://doi.org/10.1103/PhysRevD.102.065006}{\emph{Phys. Rev. D} {\bfseries 102} (2020) 065006} [\href{https://arxiv.org/abs/2001.07552}{{\ttfamily 2001.07552}}].

\bibitem{Eto:2020cys}
M.~Eto, M.~Kawaguchi, M.~Nitta and R.~Sasaki, \emph{{Exhausting all exact solutions of BPS domain wall networks in arbitrary dimensions}}, \href{https://doi.org/10.1103/PhysRevD.101.105020}{\emph{Phys. Rev. D} {\bfseries 101} (2020) 105020} [\href{https://arxiv.org/abs/2003.13520}{{\ttfamily 2003.13520}}].

\bibitem{Eto:2005cp}
M.~Eto, Y.~Isozumi, M.~Nitta, K.~Ohashi and N.~Sakai, \emph{{Webs of walls}}, \href{https://doi.org/10.1103/PhysRevD.72.085004}{\emph{Phys. Rev. D} {\bfseries 72} (2005) 085004} [\href{https://arxiv.org/abs/hep-th/0506135}{{\ttfamily hep-th/0506135}}].

\bibitem{Eto:2005fm}
M.~Eto, Y.~Isozumi, M.~Nitta, K.~Ohashi and N.~Sakai, \emph{{Non-Abelian webs of walls}}, \href{https://doi.org/10.1016/j.physletb.2005.10.017}{\emph{Phys. Lett. B} {\bfseries 632} (2006) 384} [\href{https://arxiv.org/abs/hep-th/0508241}{{\ttfamily hep-th/0508241}}].

\bibitem{Isozumi:2003rp}
Y.~Isozumi, K.~Ohashi and N.~Sakai, \emph{{Exact wall solutions in five-dimensional SUSY QED at finite coupling}}, \href{https://doi.org/10.1088/1126-6708/2003/11/060}{\emph{JHEP} {\bfseries 11} (2003) 060} [\href{https://arxiv.org/abs/hep-th/0310189}{{\ttfamily hep-th/0310189}}].

\bibitem{Isozumi:2003uh}
Y.~Isozumi, K.~Ohashi and N.~Sakai, \emph{{Massless localized vector field on a wall in D = 5 SQED with tensor multiplets}}, \href{https://doi.org/10.1088/1126-6708/2003/11/061}{\emph{JHEP} {\bfseries 11} (2003) 061} [\href{https://arxiv.org/abs/hep-th/0310130}{{\ttfamily hep-th/0310130}}].

\bibitem{Isozumi:2004jc}
Y.~Isozumi, M.~Nitta, K.~Ohashi and N.~Sakai, \emph{{Construction of non-Abelian walls and their complete moduli space}}, \href{https://doi.org/10.1103/PhysRevLett.93.161601}{\emph{Phys. Rev. Lett.} {\bfseries 93} (2004) 161601} [\href{https://arxiv.org/abs/hep-th/0404198}{{\ttfamily hep-th/0404198}}].

\bibitem{Isozumi:2004va}
Y.~Isozumi, M.~Nitta, K.~Ohashi and N.~Sakai, \emph{{Non-Abelian walls in supersymmetric gauge theories}}, \href{https://doi.org/10.1103/PhysRevD.70.125014}{\emph{Phys. Rev. D} {\bfseries 70} (2004) 125014} [\href{https://arxiv.org/abs/hep-th/0405194}{{\ttfamily hep-th/0405194}}].

\bibitem{Isozumi:2004vg}
Y.~Isozumi, M.~Nitta, K.~Ohashi and N.~Sakai, \emph{{All exact solutions of a 1/4 Bogomol'nyi-Prasad-Sommerfield equation}}, \href{https://doi.org/10.1103/PhysRevD.71.065018}{\emph{Phys. Rev. D} {\bfseries 71} (2005) 065018} [\href{https://arxiv.org/abs/hep-th/0405129}{{\ttfamily hep-th/0405129}}].

\bibitem{Eto:2004rz}
M.~Eto, Y.~Isozumi, M.~Nitta, K.~Ohashi and N.~Sakai, \emph{{Instantons in the Higgs phase}}, \href{https://doi.org/10.1103/PhysRevD.72.025011}{\emph{Phys. Rev. D} {\bfseries 72} (2005) 025011} [\href{https://arxiv.org/abs/hep-th/0412048}{{\ttfamily hep-th/0412048}}].

\bibitem{Eto:2005sw}
M.~Eto, Y.~Isozumi, M.~Nitta and K.~Ohashi, \emph{{1/2, 1/4 and 1/8 BPS equations in SUSY Yang-Mills-Higgs systems: Field theoretical brane configurations}}, \href{https://doi.org/10.1016/j.nuclphysb.2006.06.026}{\emph{Nucl. Phys. B} {\bfseries 752} (2006) 140} [\href{https://arxiv.org/abs/hep-th/0506257}{{\ttfamily hep-th/0506257}}].

\bibitem{Sakai:2006prl}
N.~Sakai, M.~Eto, Y.~Isozumi, M.~Nitta and K.~Ohashi, \emph{{Effective Lagrangians on Domain Walls and Other Solitons}}, \href{https://doi.org/10.22323/1.040.0025}{\emph{PoS} {\bfseries STRINGSLHC} (2006) 025} [\href{https://arxiv.org/abs/hep-th/0703136}{{\ttfamily hep-th/0703136}}].

\bibitem{Eto:2006pg}
M.~Eto, Y.~Isozumi, M.~Nitta, K.~Ohashi and N.~Sakai, \emph{{Solitons in the Higgs phase: The Moduli matrix approach}}, \href{https://doi.org/10.1088/0305-4470/39/26/R01}{\emph{J. Phys. A} {\bfseries 39} (2006) R315} [\href{https://arxiv.org/abs/hep-th/0602170}{{\ttfamily hep-th/0602170}}].

\bibitem{Oda:1999az}
H.~Oda, K.~Ito, M.~Naganuma and N.~Sakai, \emph{{An Exact solution of BPS domain wall junction}}, \href{https://doi.org/10.1016/S0370-2693(99)01355-6}{\emph{Phys. Lett. B} {\bfseries 471} (1999) 140} [\href{https://arxiv.org/abs/hep-th/9910095}{{\ttfamily hep-th/9910095}}].

\bibitem{Weinberg:1981eu}
E.~J. Weinberg, \emph{{Index Calculations for the Fermion-Vortex System}}, \href{https://doi.org/10.1103/PhysRevD.24.2669}{\emph{Phys. Rev. D} {\bfseries 24} (1981) 2669}.

\bibitem{Ganoulis:1987np}
N.~Ganoulis and G.~Lazarides, \emph{{A Generalized Index Theorem for String Superconductivity in Realistic Models}}, \href{https://doi.org/10.1103/PhysRevD.38.547}{\emph{Phys. Rev. D} {\bfseries 38} (1988) 547}.

\bibitem{Klimashonok:2019iya}
V.~Klimashonok, I.~Perapechka and Y.~Shnir, \emph{{Fermions on kinks revisited}}, \href{https://doi.org/10.1103/PhysRevD.100.105003}{\emph{Phys. Rev. D} {\bfseries 100} (2019) 105003} [\href{https://arxiv.org/abs/1909.12736}{{\ttfamily 1909.12736}}].

\bibitem{Weinberg:2012pjx}
E.~J. Weinberg, \emph{{Classical solutions in quantum field theory}: {Solitons and Instantons in High Energy Physics}}, Cambridge Monographs on Mathematical Physics. Cambridge University Press, 9, 2012, \href{https://doi.org/10.1017/CBO9781139017787}{10.1017/CBO9781139017787}.

\end{thebibliography}\endgroup

\end{document}